\documentclass[prd,%
preprint,%
tightenlines,superscriptaddress,showpacs,%
nofootinbib,eqsecnum,amsfonts,amsmath]{revtex4}
\usepackage{bm}

\newtheorem{Lem}{Lemma}

\def\be{\begin{equation}}
\def\ee{\end{equation}}
\def\bea{\begin{eqnarray}}
\def\eea{\end{eqnarray}}

\def\jokR{J_{0}(kR)}
\def\jokr{J_{0}(kr)}
\def\jokiR{J_{0}(k_{1}R)}
\def\jokir{J_{0}(k_{1}r)}
\def\joktR{J_{0}(k_{2}R)}
\def\joktr{J_{0}(k_{2}r)}

\def\jikr{J_{1}(kr)}
\def\jikiR{J_{1}(k_{1}R)}
\def\jikir{J_{1}(k_{1}r)}
\def\jiktR{J_{1}(k_{2}R)}
\def\jiktr{J_{1}(k_{2}r)}

\def\ki{k_{1}}
\def\kt{k_{2}}
\def\half{\frac{1}{2}}
\def\inty{\int_{0}^{\infty}}
\def\kto{k_{2(0)}}

\begin{document}

\title{Functional evolution of quantum cylindrical waves}

\author{Demian H.J. Cho}
 \email{cho@rri.res.in}
  \author{Madhavan Varadarajan} 
 \email{madhavan@rri.res.in}
 
\affiliation{
Raman Research Institute, Bangalore 560 080, India
}

\date{\today}


\begin{abstract}

\noindent
Kucha{\v r} showed that the quantum dynamics of (1 polarization) 
cylindrical wave solutions to vacuum general relativity is  
determined by that of a free axially-symmetric scalar field 
along arbitrary axially-symmetric foliations of a fixed flat 
2+1 dimensional spacetime. We investigate if such a dynamics 
can be defined {\em unitarily} within the standard Fock space 
quantization of the scalar field. 

Evolution between two arbitrary slices of an arbitrary foliation 
of the flat spacetime can be built out of a restricted class of 
evolutions (and their inverses). The restricted evolution is from 
an initial flat slice to an arbitrary (in general, curved) slice 
of the flat spacetime and can be decomposed into (i) `time' 
evolution in which the spatial Minkowskian coordinates serve as 
spatial coordinates on the initial and the final slice, followed by 
(ii) the action of a spatial diffeomorphism of the final slice on 
the data obtained from (i). We show that although the functional 
evolution of (i) is unitarily implemented in the quantum theory, 
generic spatial diffeomorphisms of (ii) are not. Our results 
imply that a Tomanaga-Schwinger type functional evolution of 
quantum cylindrical waves is not a viable concept even though, 
remarkably, the more limited notion of functional evolution 
in Kucha{\v r}'s `half parametrized formalism' is well-defined.

\end{abstract}
\maketitle

\widetext

\section{Introduction}

Given the difficulties of constructing a quantum theory of gravity, 
it is of use to build intuition from the study of simpler toy models. 
While cosmological mini-superspaces \cite{misner,bojowald} have 
yielded valuable insights, their finite-dimensional nature precludes 
the occurrence of field-theoretic aspects of quantum gravity. 
In this regard, vacuum cylindrically symmetric gravitational fields 
with 2 hypersurface-orthogonal, commuting Killing vectors (one generating 
translations along `$Z$'  and the other, rotations along `$\Phi$') 
constitute a useful infinite-dimensional midi-superspace. Quantization 
of this cylindrical wave midi-superspace was initiated by Kucha{\v r}
\cite{karelcylwave} and further studied by a number of authors 
\cite{ashtekarpierri,me,allen,barbero,mena}. While the latter works 
deal with quantization after fixing a gauge, our focus here is on 
aspects of Dirac quantization of cylindrical waves.

Every cylindrical wave solution is determined by a corresponding 
axially-symmetric solution to the free scalar wave equation on a 
fixed 3 dimensional flat spacetime. Specifically, the cylindrical 
wave line element is
\be
ds^{2} = e^{\gamma - \psi} ( -(dT)^{2} + (dR)^{2} ) 
+ R^{2}e^{- \psi}( d\Phi )^{2} + e^{ \psi}(dZ)^{2}
\label{cylmetric}
\ee
and that of the flat spacetime is
\be
ds^{2} =   -(dT)^{2} + (dR)^{2}  + R^{2}( d\Phi )^{2}.
\label{flatmetric}
\ee
$\psi (R,T) $ is the free scalar field and $\gamma (R,T)$ is 
its energy in a box of size $R$ at time $T$. Kucha{\v r} defined 
a canonical transformation from the Arnowitt-Deser-Misner (ADM) 
phase space of cylindrical waves to that of an axially-symmetric 
parametrized field theory on 2+1 dimensional flat spacetime 
and studied the formal Dirac quantization of the latter system 
\cite{karelcylwave}.

Parametrized field theory (PFT) is free field theory on flat 
spacetime in a diffeomorphism invariant disguise \cite{kareliyer}. 
It describes field evolution on {\em arbitrary} (and in general, curved) 
foliations of the flat spacetime instead of only the usual flat 
foliations, by treating the `embedding variables' which describe 
the foliation as dynamical variables to be varied in the action 
in addition to the scalar field. In the present context the coordinates 
$X^{\alpha}:= (T,R)$ are parametrized by a new set of arbitrary 
coordinates $x^{\alpha}=(t,r)$ such that  for fixed  $t$, 
$X^{\alpha}(t,r)$ define an axisymmetric spacelike slice in 
Minkowski spacetime with radial coordinate $r$. 
General covariance of PFT ensues from the arbitrary choice of 
$x^{\alpha}$. In the Hamiltonian description, the embedding 
variables `$X^{\alpha}$' are canonical coordinates and general 
covariance implies that Hamiltonian evolution from one slice of 
an axisymmetric foliation to another is generated by constraints. 

Indeed, under Kucha{\v r}'s canonical transformation \footnote
{While $R$ is directly related to the radial components of the 
metric (\ref{cylmetric}), Kucha{\v r}'s canonical transformation 
identifies $T$ with the spatial integral of one of the extrinsic 
curvature components of the cylindrical wave metric (\ref{cylmetric}) 
on a $t$= constant slice.}, appropriate combinations of 
the Hamiltonian and diffeomorphism constraints of the midi-superspace 
take the form of the constraints appropriate to PFT, namely,
\be
{C}_{\alpha}(x) := P_{\alpha}(x) 
+ h_{\alpha}[\psi,\pi,X^{\alpha}](x) \approx 0 \;\;,
\label{calpha}
\ee
where $P_{\alpha}$ and $\pi$ are the momenta conjugate to 
$X^{\alpha}$ and $\psi$, and $h_{\alpha}$ is related to the 
stress-energy of the scalar field \footnote{As is standard in 
constraint dynamics, $\approx$ denotes weak equality.}. 
The gauge fixing conditions $X^{\alpha} = x^{\alpha}$ \footnote{
Technically, this is not a gauge fixing but a 1-parameter 
family of gauge fixings (one for each value of $t$) known as 
`deparameterization'.} map the classical theory directly onto 
the standard description of a free scalar field in Minkowskian 
coordinates $(R,T)$. This description admits the standard Heisenberg 
picture based Fock quantization\cite{karelcylwave}\cite{ashtekarpierri}. 

In the Dirac quantization, a formal operator version of 
$C_{\alpha}$ acting on a physical state $\arrowvert \Psi \rangle$ 
of the theory is given by
\be
\left(\frac{1}{i}\frac{\delta}{\delta X^{\alpha}} 
+ \hat{h}_{\alpha} \right) \arrowvert \Psi \rangle = 0 \;\;.
\label{fsch}
\ee
Eq.(\ref{fsch}) takes the form of a functional Schr\"odinger 
equation which represents infinitesimal evolution of the Schr\"odinger 
state $\arrowvert \Psi \rangle$ from one Cauchy slice to another. 
In this work we investigate if the formal equation (\ref{fsch}) 
is well defined in the context of the standard Heisenberg picture 
based Fock space quantization of the free scalar field on flat spacetime. 
More precisely, in analogy with the relation between the Heisenberg and 
Schr\"odinger pictures in quantum mechanics, we investigate if there 
exists a unitary transformation on the standard Fock space which implements 
evolution between any two axisymmetric Cauchy slices of the 
flat spacetime and whose infinitesimal version yields equation (\ref{fsch}). 

What do we mean by evolution between two slices? Since the classical 
theory is that of a free scalar field we explicitly (and uniquely) 
know the evolved data $(\psi, \pi)$ on the slice $t=t_1$ in terms of 
initial data at $t=t_0$. Since the equations for scalar field evolution 
are linear, we also know how the corresponding quantum operators are 
related. This is what we mean by evolution. Since this notion of 
evolution is independent of the choice of interpolating foliation 
between the two slices, our subsequent considerations are phrased 
solely in terms of data on the two slices.

We enquire if the operators at the initial and final slices are 
unitarily related. If they are, then in analogy to the definition 
of the Schr\"odinger picture from the Heisenberg picture in usual 
quantum mechanics, we can define the Schr\"odinger state at any time 
$t$ as the unitary image of the state at time $t_0$ 
\footnote{For a beautiful exposition of the Schr\"odinger 
and Heisenberg pictures in a canonical treatment of PFT on 
(n+1)-dimensional flat spacetime (in the absence of axisymmetry), 
we refer the reader to \cite{kareliyer,karel1+1}.} 
\footnote{ Note that if we restrict attention to only flat slices, 
the standard Hamiltonian is the generator of the desired unitary 
transformation. The non-triviality enters solely due to the possibility 
of evolution along directions which are not isometries of the spacetime.}. 
Techniques appropriate to our investigation here have been developed 
in \cite{charlie1,charlie2} for the case of PFT on (n+1)-dimensional 
flat spacetime in the absence of axisymmetry. That work arose as 
an effort to rigorously implement the considerations of 
Kucha{\v r} in \cite{kareliyer,karel1+1}. The strategy pursued in 
\cite{charlie1,charlie2} is as follows. Classical scalar field 
evolution from an initial flat slice to any final slice is 
a linear canonical transformation on the scalar field phase space. 
Instead of phrasing this canonical transformation in terms of 
the scalar field and its momentum, one can instead use 
appropriate linear functionals of these fields. The linear 
functionals used are such that when they are evaluated on the 
initial slice, they reduce to the familiar mode coefficients 
(i.e. the classical correspondents of the standard annihilation 
and creation operators). It then follows (from the fact that evolution 
is a canonical transformation) that these functionals evaluated 
on the final slice are related by a Bogolubov transformation 
to their initial values. The criterion for unitary implementability 
of a Bogolubov transformation is that the `$\beta$' matrix 
be Hilbert-Schmidt (i.e. the squares of the  absolute values of 
its diagonal elements should be summable) \cite{shale,wald}.

The results of \cite{charlie1,charlie2} are as follows. 
In 1+1 dimensions operator evolution is unitary, the functional 
Schr\"odinger picture exists as the unitary image of the standard 
Heisenberg-picture-based Fock quantization and the infinitesimal 
version of the unitary transformation on Heisenberg states implies 
that the Schr\"odinger states satisfy an equation of the form 
(\ref{fsch}) (albeit, with an extra `anomaly potential' term 
\cite{karel1+1,charlie1}). In $n+1$ dimensions ($n>1$), 
the $\beta$ coefficients for evolution along generic 
curved-foliations are not Hilbert-Schmidt. Hence the functional 
Schr\"odinger picture does not exist (at least in the form 
usually envisioned, as the unitary image of the Heisenberg picture) 
and the functional Schr\"odinger equation cannot be given 
any obvious meaning. The case of interest here, namely 2+1 dimensions 
with fewer degrees of freedom (only the axisymmetric ones), 
is a tantalizing intermediate case and as we shall demonstrate, 
we get a suitably `in between' answer.

Here, our interest is in evolution between two axisymmetric, 
but otherwise arbitrary, Cauchy slices $\Sigma_i$ and $\Sigma_f$ 
in the flat spacetime (\ref{flatmetric}). Note that the restriction of 
the spacetime coordinate $R$ to any axisymmetric spatial slice 
defines a natural radial coordinate on the slice. We refer to 
this natural coordinate by the same symbol $R$. 
In order to accommodate arbitrary reparameterizations 
$(R,T) \rightarrow (r,t)$, it is necessary to admit arbitrary
radial coordinates $r_f (R)$ on $\Sigma_f$ and $r_i(R)$ on $\Sigma_i$
\footnote{Here $\Sigma_{i}$ and $\Sigma_{f}$ are slices of 
constant $t$, i.e. $t=t_{i}$ and $t=t_{f}$. Hence $r$ is the radial 
coordinate on $\Sigma_{i}$ and $\Sigma_{f}$. Since $r$ is in general 
different from the coordinate $R$, $r=r_{i}(R)$ and $r=r_{f}(R)$ 
express the functional relation between the two coordinates on 
$\Sigma_{i}$ and on $\Sigma_{f}$.}. 
Despite this, it suffices for our purposes, to consider evolution 
from a {\em fixed} initial slice $\Sigma_0$, chosen to be the $T=0$ 
slice coordinatized by the natural radial coordinate $R$, 
to an arbitrary final slice $\Sigma$, coordinatized by an arbitrary 
radial coordinate $r=r(R)$. This entails no loss of generality since, 
if in the quantum theory all evolution is unitary, quantum evolution 
from any $\Sigma_i$ to any $\Sigma_f$ may be constructed as 
the (inverse) unitary evolution from $\Sigma_i$ to $\Sigma_0$ 
followed by unitary evolution from $\Sigma_0$ to $\Sigma_f$.

Now, consider the restricted evolution from $\Sigma_0$ to $\Sigma$ 
such that the radial coordinate $r$ on $\Sigma$ is chosen to be 
the natural radial coordinate $R$ i.e. $r$ is identified 
with the restriction of the spacetime coordinate $R$ to $\Sigma$. 
We shall refer to such evolution as the `half parametrized' evolution
\cite{karelcylwave}. Next, we reconsider the general case of interest, 
namely that of evolution from $\Sigma_0$ (with coordinate $R$) to 
$\Sigma$ with coordinate $r=r(R)$. The uniqueness of evolution from 
initial data implies that $(\psi (r), \pi (r))$ on $\Sigma$ can be 
obtained by first evolving the initial data from $\Sigma_0$ to 
$\Sigma$ by a `half parametrized' evolution and then subjecting 
the result to the spatial diffeomorphism defined by $r(R)$.

In view of this we proceed as follows. In section II we construct 
the Bogolubov transformation corresponding to the most general 
evolution from $\Sigma_0$ to $\Sigma$. In section III, we particularize 
the Bogolubov transformation of section II to the case of 
`half parametrized' evolution defined above and show that the 
$\beta$ matrix is Hilbert-Schmidt. This implies that the finite 
functional evolution corresponding to Kucha{\v r}'s `half parameterized' 
formalism \cite{karelcylwave} is well defined. In section IV we show that 
the action of a generic spatial diffeomorphism on data on any slice 
$\Sigma$ is not unitarily implemented in quantum theory. This implies 
that the most general Tomonaga- Schwinger type functional evolution 
of quantum cylindrical waves is not a viable concept. In section V 
we digress from the main theme of this paper and turn our attention 
to the generators of spatial diffeomorphisms of $\Sigma_0$. 
Specifically, we show that such generators generically do not have 
a well-defined action on the Fock vacuum. Section VI 
contains a discussion of our results and of open issues. Some 
technicalities are collected in the Appendices. 

In what follows we 
set $\hbar=c=1$. Note that in our discussion of the relation between 
cylindrical waves and PFT, Newton's constant ``per unit $Z$''
\cite{ashtekarpierri} has been set to $\frac{1}{8}$.

\vspace{0.3 in}
\noindent
{\bf Note on notations}
\vspace{0.1 in}

The quantum counterpart of a classical quantity, $X$, is denoted by 
$\hat{X}$, its complex conjugate is denoted by $X^{*}$. 
$\partial _{t} X$,  $\partial _{r} X$ are denoted by $\dot{X}$ and 
$X^{\prime}$, respectively. The adjoint of $\hat{X}$ is 
${\hat{X}^{\dagger}}$ and its normal-ordered version is $: \hat{X} :$.
$f(y,k) = O(1/k^{n})$ means that, for $k \rightarrow \infty$, 
$\exists$ a positive number $C$ such that $|f(y,k)| \leq C/k^{n}$, 
with $C$ independent of $y$.

\section{Evolution as a Bogolubov transformation}
\label{sec:review}

The axisymmetric free scalar field, $\psi (R, T) $, satisfies 
the wave equation on the flat spacetime of equation (\ref{flatmetric}).
Hence, we have that
\be
-\frac{\partial^{2}\psi}{\partial T^{2}} 
+ \frac{\partial^{2}\psi}{\partial R^{2}} 
+ \frac{1}{R}\frac{\partial\psi}{\partial R} = 0\;\;.
\ee

The mode expansion of $\phi$ which is regular at the axis is
\be
\psi(R,T) = \frac{1}{\sqrt{2}}\int_{0}^{\infty} dk \; \jokR \; 
[\;a(k)e^{-ikT} + a^{*}(k)e^{ikT}\;]\;\;.
\label{psi}
\ee

Consider an arbitrary axisymmetric foliation of the flat spacetime 
(\ref{flatmetric}) coordinatized by $(t,r)$ (each leaf of the foliation 
is at constant $t$; $r$ is the radial coordinate on each leaf). 
Let $\Sigma$ be some slice at time $t$. The momentum, $\pi (r,t)$, 
(canonically conjugate to the scalar field $\psi (r,t)$ on this slice) 
obtained through a Hamiltonian decomposition of the free scalar field 
action with respect to the foliation \cite{karelcylwave} is
\begin{equation}
\pi = \frac{R}{({\dot T} R^{\prime} 
- {\dot R} T^{\prime})}[(R^{\prime 2}-T^{\prime 2})\dot{\psi} 
- ({\dot R}R^{\prime}-{\dot T}T^{\prime})\psi^{\prime}]\;\;.
\label{pirt}
\end{equation}
Here $(R=R(t,r),T= T(t,r))$ denote the spacetime coordinates $(R,T)$ 
evaluated at the point $(r,t)$ on the foliation. As is conventional 
in the Hamiltonian description, we shall suppress the label $t$ and 
denote the scalar field and its momentum on $\Sigma$ by 
$(\psi (r), \pi (r))$ and the restriction of the spacetime coordinates 
$R,T$ to $\Sigma$ by $R(r), T(r)$. A useful form of $\pi (r)$ for our 
subsequent considerations is obtained by re-expressing the $r,t$ 
derivatives of $\psi$ in Eq. (\ref{pirt}) in terms of $R,T$ derivatives 
so that
\be
\pi (r) = R(R' \psi ,_{T} + T' \psi ,_{R} ) \;\;,
\label{momenta}
\ee
where $\psi ,_{T}$ and $\psi ,_{R}$ denote derivatives of $\psi$ with 
respect to $T$ and $R$ respectively.

Next, we trade $(\psi (r), \pi (r)) $ for suitable linear functionals 
thereof. We require that these functionals reduce to the mode coefficients 
$a(k)$ of Eq. (\ref{psi}) when $\Sigma$ is chosen to be the flat 
slice $\Sigma_0$ at $T=0$ with radial coordinate $r=R$. Note that from 
Eq.(\ref{psi}) and Eq.(\ref{momenta}), the field and the momenta on 
$\Sigma_0$, $\psi_{0}$ and $\pi_{0}$, are
\bea
\psi_{0} &=& \frac{1}{\sqrt{2}}\int_{0}^{\infty} dk \; \jokr \; 
[\;a(k) + a^{*}(k)\;]\;,\; \nonumber\\
\pi_{0} &=& -\frac{i}{\sqrt{2}} r \int_{0}^{\infty} dk \; k \; \jokr \; 
[\;a(k) - a^{*}(k)\;]\;.
\label{zero}
\eea
Using the orthogonality of Bessel functions we can invert 
Eq.(\ref{zero}) to get 
\be
a(k) = \frac{1}{\sqrt{2}}\int_{0}^{\infty} dr 
\left[ i \; \jokr \pi_{0}(r) - r \; \jikr \; \psi^{\prime}_{0}(r)\; \right]\;.
\label{ak}
\ee
The evolution from $\Sigma_0$ to $\Sigma$ of $a(k) = a(k)[\psi_0, \pi_0]$ 
(where the right hand side of the equality represents the functional of 
$\psi_0, \pi_0$ appearing on the right hand side of Eq.(\ref{ak})) 
is obtained by replacing the fields $\psi_0, \pi_0$ by their evolved 
images $\psi (r), \pi (r)$ on $\Sigma$. Thus we have that (in obvious notation)
\be
a_{\Sigma}(k) = \frac{1}{\sqrt{2}}\int_{0}^{\infty} dr 
\left[ i \; \jokr \pi(r) - r \; \jikr \; \psi^{\prime}(r)\; \right]\;\;. 
\label{ck}
\ee
Using Eq.(\ref{psi}) and Eq.(\ref{momenta}), $\psi (r) , \pi (r) $ can be 
re-expressed in terms of $a(k), a^*(k)$ to obtain an equation of the form
\be
a_{\Sigma}(\ki) = \inty d\kt \left[ \alpha(\ki,\kt)\;a(\kt) 
+ \beta(\ki,\kt)\;a^{*}(\kt) \right] \;\;.
\label{bogo}
\ee
Although we do not display the calculations here, it can be checked that 
$\alpha$ and $\beta$ satisfy the standard Bogolubov conditions 
(see for example \cite{birelldavis}). On general grounds this is expected 
since, as mentioned earlier, evolution is a canonical transformation.
$\beta(\ki,\kt)$ is given by 
\bea
\beta(\ki,\kt) &=&\frac{\kt}{2} \int_{0}^{\infty} dr\;r 
                  \left(\left[\jikir\jiktR - (\frac{R}{r})
                  \jokir\joktR\right]R'\right.\nonumber\\ 
               &-&\left.i\left[(\frac{R}{r})\jokir\jiktR 
                  + \jikir\joktR\right]\;T' \right) e^{i\kt T} \;\;.
\label{betagen}
\eea
Since we do not need the expression for $\alpha$ in this work, 
we do not display it here.

From the work of Shale \cite{shale}, the canonical transformation defined 
by Eq. (\ref{bogo}) is unitarily implemented iff $\beta$ satisfies the 
Hilbert- Schmidt condition 
\be
\int_{0}^{\infty} d\ki\;\int_{0}^{\infty} d\kt\; |\beta|^2\;<\;\infty\;\; .
\label{hilbert}
\ee

We end this section with a specification of the class of 
embeddings $(T(r), R(r))$ for which we analyse (\ref{hilbert}). 
We require that the embeddings be `Minkowskian' near  the axis of 
symmetry, $r=0$, and at spatial infinity. 
Specifically, we require that $R(r) -r$ and $T^{\prime}(r)$ be 
compactly supported away from $r=0$ and $r\rightarrow \infty$. 
Hence, on every slice $\Sigma$ we require that  there exist positive 
$a,b,c,d$  with $b>a$, $d>c$ such that
\be
T^{\prime}(r) = 0 \;\;\;{\rm for} \;r \in [0,a) \; \cup \;(b, \infty)\;\;,
\label{bct}
\ee
\be
R(r) -r = 0 \;\;\; {\rm for} \; r \in [0,c) \; \cup \; (d, \infty)\;\;.
\label{bcr}
\ee
Note that (\ref{bct}) and (\ref{bcr}) ensure that the slice $\Sigma$  
does not have a `cusp' at the axis.

For future reference, we note that the spatial nature of  $\Sigma$  
implies that
\be
R^{\prime 2}- T^{\prime 2} >0\;\;.
\label{nondeg1}
\ee
In particular, the above equation in conjunction with (\ref{bcr}) 
implies that
\be
R^{\prime} > 0\;\;,
\label{nondeg2}
\ee
so that $R$ can indeed be used as a radial coordinate on $\Sigma$.
Finally, we shall also assume that $R, T$ are $C^{\infty}$-functions 
of $r$.
\section{ The `half parametrized' evolution}
In this section we show that the `half parametrized' evolution 
satisfies the Hilbert-Schmidt condition, Eq.(\ref{hilbert}). 
As mentioned in the introduction, `half parametrized' evolution 
is defined as evolution in which the radial coordinate `$r$' on 
$\Sigma$ is chosen to be the restriction of the spacetime 
coordinate `$R$' to $\Sigma$. Setting $R = r$ in Eq.(\ref{betagen}), 
we get 
\bea
\beta(\ki,\kt) &=&\frac{\kt}{2} \int_{0}^{\infty} dr\;r ([\jikir\jiktr 
                  - \jokir\joktr]\nonumber\\  
               &-& i[\jokir\jiktr + \jikir\joktr]\;T' )e^{i\kt T} \;\;.
\label{bt}
\eea
For later calculations, however, a more symmetric form of $\beta$ 
turns out to be convenient. As shown in Appendix \ref{sec:newapp}, 
using Bessel function identities, 
$\frac{d}{dz}\left [ z^{n}J_{n}(z) \right] = z^{n}J_{n-1}(z)$, 
$\frac{d}{dz} J_{0}(z) = - J_{1}(z)$, their orthogonality relations, and 
integration by parts in conjunction with Eq.(\ref{bct}), we get
\be
\beta(\ki,\kt) = -\frac{i}{2}\frac{\ki\kt}{(\ki + \kt)} 
                  \int_{a}^{b} dr\;r [\jokir\jiktr 
                  +\jikir\joktr]\;T' e^{i\kt T} \;\;.
\label{btt}
\ee

Clearly, divergent contributions to Eq.(\ref{hilbert}) may be 
obtained only when $\ki$ or $\kt$ or both diverge. We estimate these 
`ultra-violet' contributions to the integral in (\ref{hilbert}) 
by considering the following four exhaustive cases.

\vspace{0.2in}
\noindent
A.1. $\kt \rightarrow \infty,\;\; 0 < \ki < \kt^{1/3}$.\\
A.2. $\ki \rightarrow \infty,\;\; 0 < \kt < \ki^{1/3}$.\\
B.1. $\kt,\ki \rightarrow \infty,\;\; \kt^{1/3} \leq \ki \leq \kt$.\\
B.2. $\ki,\kt \rightarrow \infty,\;\; \ki^{1/3} \leq \kt \leq \ki$.

\vspace{0.2in}
\noindent

\vspace{0.5in}
\noindent
[Case A.1] $\kt \rightarrow \infty;\;\; 0 < \ki < \kt^{1/3};\;\;\kt a \gg 1.$\\
Since $\kt a \gg 1$, we use the asymptotic expansion of Bessel functions, 
Appendix \ref{sec:asymptotic}, for the ones with $\kt r$ as their argument. 
The integral contributing to the first term in Eq.(\ref{btt}) is
\bea
\int_{a}^{b} &dr&\;r\jokir\jiktr\;T^{\prime} e^{i\kt T} \nonumber\\ 
             &=& \frac{1}{\sqrt{2\pi\kt}}\int_{a}^{b} dr\;
                 \sqrt{r}T^{\prime}\jokir\{ e^{-i 3\pi/4}e^{i\kt (r+T)} 
                 + e^{i 3\pi/4}e^{-i\kt (r-T)} 
                 + O\left(\frac{1}{\kt}\right)\}\;\;.\label{caseaa}
\eea
Since $T^{\prime}$ has compact support away from the origin, 
$\sqrt{r}T^{\prime}\jokir$ is a smooth, bounded function of 
compact support. Further, for space-like embeddings, $(r \pm T)' \neq 0$ 
because $|T'| < 1$ (this follows from Eq.(\ref{nondeg1}) with $R = r$). 
Then the integral in Eq.(\ref{caseaa}) is a linear combination of  
integrals of the form considered in Lemma 1 with $f_{3} = 0$. 
A similar analysis yields the same conclusion for the second term in 
Eq.(\ref{btt}). Hence, using Lemma 1, $\beta$ can be estimated as
\be
|\beta(\ki,\kt)| < \frac{\ki\kt}{(\ki + \kt)}
                   \left[\left(\frac{\ki}{\kt^{3/2}}M_{1} + 
                   \frac{1}{\kt^{3/2}}M_{2}\right) \right] \;\;,
\label{betaa}
\ee
for suitable positive real numbers $M_{1}, M_{2}$.   
It follows that
\be
|\beta(\ki,\kt)|^2 < \frac{\ki^{2}}{\kt^{3}}\left[ C_{1}  
+ C_{2} \ki + C_{3} \ki^{2} \right]\;\;,
\ee
for suitable $C_{1}, C_{2}, C_{3}$.
This gives the estimate
\be
\int_{0}^{\kt^{1/3}}d\ki |\beta(\ki,\kt)|^2 < \frac{C_{1}}{3}\frac{1}{\kt^{2}} 
+ \frac{C_{2}}{4}\frac{1}{\kt^{5/3}} + \frac{C_{3}}{5}\frac{1}{\kt^{4/3}}\;\;,
\ee
which in turn implies convergence of the integral in Eq.(\ref{hilbert}) 
in the domain of integration considered here.

\vspace{0.5in}
\noindent
[Case A.2]  $\ki \rightarrow \infty;\;\;0 < \kt < \ki^{1/3};\;\;\ki a \gg 1$.\\
The only non-symmetric part (with respect to $\ki$ and $\kt$) of 
$\beta(\ki,\kt)$ in Eq.(\ref{btt}) is $e^{i\kt T}$, which results 
in the terms $e^{i(\kt T \pm \ki r)}$ rather than $e^{i\kt (r \pm T)}$ 
in Eq.(\ref{caseaa}). However, Lemma 1 still applies. The rest of 
the argument is similar to the one for [Case A.1], and also yields 
a convergent contribution to Eq.(\ref{hilbert}) in this domain of 
integration.

\vspace{0.5in}
\noindent
[Case B.1]  $\kt,\ki \rightarrow \infty;\;\;
\kt^{1/3} \leq \ki \leq \kt;\;\; \ki a, \kt a  \gg 1.$\\
In this case we can expand all the Bessel functions using their 
asymptotic forms for large arguments. Then we have that, 
\bea
\beta(\ki,\kt) &=& \frac{i}{8\pi}\frac{(\ki\kt)^{1/2}}{(\ki + \kt)} 
                   \int_{a}^{b}dr\Biggl[ 8\cos (\ki + \kt)r
                   - [\sin (\ki + \kt)r \Biggr. \nonumber\\
               &+& \Biggl. 2\cos (\ki -\kt)r ]\left(\frac{1}{\ki} 
                   + \frac{1}{\kt}\right)\frac{1}{r} 
                   + O\left(\frac{1}{\ki^2}\right)\Biggr]T'e^{i\kt T} 
                   \label{caseb1} \\
               &:=&\beta_{+} + \beta_{-} + K\;\;. \label{caseb1b}
\eea
Here $\beta_{+}$ is the sum of the first two terms (i.e. the 
$\cos (\ki+\kt)r$ and the $\sin (\ki+\kt)r$ terms), while $\beta_{-}$ 
is the third term (with the $\cos (\ki-\kt)r$). $K$ denotes the remainder 
of order $1/\ki^2$. 

First we show that $\beta_{+}$ falls off faster than any power of 
$\frac{1}{(\ki + \kt)}$. The two terms in $\beta_{+}$ are of the form
\be
f_{\pm}(\ki, \kt)\int_{a}^{b} dr g_{\pm}(r)[e^{i(\ki + \kt)r}  
          \pm e^{-i(\ki + \kt)r}]T'e^{i\kt T} \;\;,
\label{z1}
\ee
where $g_{\pm}(r), T(r)$ are smooth bounded 
functions of $r$ in $[a,b]$. Consider the first 
term in (\ref{z1}). Define $\chi (r) 
:= T(r) + r$, $t := \frac{\ki}{\ki + \kt}$ and $\chi_{t}$ as 
\be
\chi_{t} := t r + (1-t) \chi (r) \,\,.
\ee
Since $|T'| < 1$ for a space-like embedding, $\chi_{t}^{'} \neq 0$ 
in the domain of the integration. This implies that there exists a 
$C^{\infty}$ function $\psi_{t} (y) = r$ where $\chi_{t} (r) = y$.  
Then,
\be
\int_{a}^{b} dr g_{\pm}(r)e^{i(\ki + \kt)r}T'e^{i\kt T} 
= \int_{\chi_{t}(a)}^{\chi_{t}(b)}dy g_{\pm}(r(y))
\frac{d\psi_{t}}{dy} T^{'}(r(y))e^{iy(\ki + \kt)} \,\,.
\label{z3}
\ee
This is an integral of the form in Lemma 2, $\bf{L.2.a}$. Therefore 
the integral falls off faster than any power of $\frac{1}{(\ki + \kt)}$ in 
the large $\ki$ and $\kt$ limit. Similar arguments apply to the second term 
in (\ref{z1}), so we conclude that $\beta_{+}$ falls off faster 
than any power of $\frac{1}{(\ki + \kt)}$ in the large $\ki$, $\kt$ limit.

Next, consider $K$. Since $\kt + \ki > \kt$ and $T^{\prime}$ is 
bounded
\be
\left|K\right| = \left| \frac{i}{8\pi}\frac{(\ki\kt)^{1/2}}{(\ki + \kt)} 
\int_{a}^{b}dr T^{\prime}e^{i\kt T} O\left(\frac{1}{\ki^{2}}\right) \right| < 
\frac{C_{4}}{\kt^{1/2} \ki^{3/2}} 
\label{R}
\ee
with a suitable positive $C_{4}$.     

Finally, consider $\beta_{-}$, 
\be
\beta_{-} = -\frac{i}{4\pi (\ki\kt)^{1/2}} \int_{a}^{b}dr
                 \cos ((\ki -\kt)r) \frac{T^{\prime}}{r}e^{i\kt T} \;\; .
\label{beta5}
\ee
Since the integrand is bounded, a rough estimate for $\beta_{-}$ yields 
$|\beta_{-}| = O((\ki\kt)^{-1/2})$. From Eq.(\ref{caseb1b})
\be
|\beta|^{2} \leq |\beta_{+}|^{2} + |\beta_{-}|^{2} + |K|^{2} 
+ 2|\beta_{+}\beta_{-}| + 2|\beta_{+} K| + 2|\beta_{-} K|\;\;.
\label{betasq}
\ee
Since $|\beta_{+}|$ falls off faster than any power of $\frac{1}
{(\ki+\kt)}$, so do $|\beta_{+}|^{2}, |\beta_{+}\beta_{-}| $, 
and $|\beta_{+} K| $. Hence these terms  give convergent contributions 
to the integral in Eq.(\ref{hilbert}).  

Next, from Eq.(\ref{R}), and $\beta_{-} = O((\ki\kt)^{-1/2})$, 
we estimate the contributions from 
$|K|^{2}$ and $2|\beta_{-} K|$ as
\be
\int_{\kt^{1/3}}^{\kt} \; d\ki \; |K|^{2} 
< \frac{C_{4}^{2}}{\kt^{5/3}} \;\;,
\label{R2}
\ee
and
\be
\int_{\kt^{1/3}}^{\kt} \; d\ki \;  |\beta_{-} K| 
< \frac{C_{5}}{\kt^{4/3}}\;\;,
\label{Rbeta}
\ee
for some appropriate positive constant $C_{5}$, which lead to 
convergent contributions to the integral in Eq.(\ref{hilbert}).

So far all the contributions to Eq.(\ref{hilbert}) except 
the one from $\left|\beta_{-}\right|^2$ have been shown to be 
convergent. The rough estimate of $\beta_{-} = 
O\left((\ki\kt)^{-1/2}\right)$, used above, however, seems to 
indicate that this contribution diverges. But, as we shall show now, 
a better estimate of $\beta_{-}$ yields a convergent contribution.

We introduce an auxiliary variable $\alpha := \frac{\kt - \ki}{\kt}$. 
Clearly, $0\leq \alpha <1$. $\beta_{-}$ is then given by
\be
\beta_{-} =  -\frac{i}{8\pi (\ki\kt)^{1/2}} \int_{a}^{b}dr
             \frac{T^{\prime}}{r} \left[ e^{i\kt ( T -\alpha r )} 
             + e^{i\kt ( T +\alpha r )}\right] \;\; .
\label{betaminus}
\ee
Since $|T^{\prime}| < 1$ and $[a,b]$ is compact, there exists a number 
$\bar{m}$ such that $|T^{\prime}| \leq \bar{m} < 1$. Furthermore, let 
$m := \bar{m} + \frac{1-\bar{m}}{2},\; m > \bar{m}$. Consider 
Eq.(\ref{betaminus}) when $\alpha > m$. If we naively estimate the 
integral in Eq.(\ref{betaminus}) as of $O(1)$, the prefactor, 
$(\ki\kt)^{-1/2} = O(1/\kt^{2/3})$, gives divergent contribution to 
Eq.(\ref{hilbert}). However, since $T^{\prime} \pm \alpha \neq 0$, 
the integral falls off faster than any power of $1/\kt$ by Lemma 2 
$\bf{L. 2. a}$. Thus, both terms in Eq.(\ref{betaminus}) give convergent 
contributions. 

The remaining contribution to be estimated is when $\alpha \leq m$ 
(That is, $(1-m)\kt \leq \ki$). We obtain, for an appropriate constant 
$ C,\; C > 0$, for $0 \leq \alpha \leq m$, 
\be
\left| \beta_{-} \right| ^{2} \leq
\frac{C}{\kt ^{2}} \int_{a}^{b} \int_{a}^{b} dr_{1} dr_{2}
\frac{T^{\prime}(r_{1})T^{\prime}(r_{2})}{r_{1}r_{2}} e^{i\kt 
\left[ T(r_{1}) - T(r_{2}) \right]}
\left[ \cos \alpha \kt (r_{1} + r_{2}) 
+ \cos \alpha \kt (r_{1} - r_{2}) \right]\;\;.
\label{longeq}
\ee
Since the prefactor is $O(\kt^{-2})$, we still need the tiniest 
power of $1/\kt$ from the integral involving $\cos \alpha \kt (r_{1} 
\pm r_{2})$ for convergence. The key point is to change variables 
from $(\ki, \kt)$ to $(\alpha, \kt)$ and integrate 
$\left| \beta_{-} \right| ^{2}$ over $\alpha$. The integral of 
$\cos\kt\alpha(r_{1} + r_{2})$ over $\alpha$ gives a factor of 
$1/\kt$ since $(r_{1} + r_{2}) \geq a$. Since $d\ki d\kt = 
\kt d\kt d\alpha$, this term contributes 
$\int \kt d\kt O\left(\kt^{-3}\right)$ which is convergent. 

The only remaining non-trivial contribution is from the term with 
$\cos \alpha \kt (r_{1} - r_{2})$. From Eq.(\ref{longeq}) this 
contribution can be estimated as
\bea
\frac{C}{\kt ^{2}}\int_{0}^{m} &d\alpha& 
\int_{a}^{b} \int_{a}^{b} dr_{1} dr_{2}
 \frac{T^{\prime}(r_{1})T^{\prime}(r_{2})}{r_{1}r_{2}} e^{i\kt 
    \left[ T(r_{1}) - T(r_{2}) \right]}
    \cos \alpha \kt (r_{1} - r_{2})\nonumber\\
&\leq& \left| \frac{C}{\kt^{2}} \int_{a}^{b} \int_{a}^{b} dr_{1} dr_{2}
       \frac{ T^{\prime}(r_{1})T^{\prime}(r_{2})}
        {r_{1}r_{2}}e^{i\kt 
    \left[ T(r_{1}) - T(r_{2}) \right]} 
       \int_{0}^{m} d\alpha \cos \alpha \kt (r_{1} - r_{2}) 
       \right| \;\;\nonumber \\
&\leq&    \frac{C}{\kt^{2}} \int dr_{+} dr_{-} f(r_{-},r_{+}) \left| 
       \frac{\sin \kt r_{-} m}{\kt r_{-}} \right| \nonumber \\
&=&    \frac{C}{\kt^{2}} \int dr_{+} \int _{r_{-} > 0} dr_{-} 
       f(r_{-},r_{+}) \left| 
       \frac{\sin \kt r_{-} m}{\kt r_{-}} \right|\nonumber\\
&&     \;\;\;+\frac{C}{\kt^{2}} \int dr_{+} \int _{r_{-} < 0} dr_{-} 
       f(r_{-},r_{+}) \left| 
       \frac{\sin \kt r_{-} m}{\kt r_{-}} \right| \;\;\nonumber \\
&:= & I_{I} + I_{II} \;\;.
\label{tanhueser}
\eea
where $r_{-} := r_{1} - r_{2}, r_{+} := r_{1} + r_{2}$ and $f(r_{-}, r_{+}) 
;= \left|\frac{ T^{\prime}(r_{1})T^{\prime}(r_{2})}{2r_{1}r_{2}}\right|$ 
is a positive function of compact support in $(r_{1}, r_{2})$, 
and hence in $(r_{-}, r_{+})$ plane. Now we estimate $I_{I}$, 
\bea
I_{I} &=& \frac{C}{\kt^{2}} \int dr_{+} \left[m \int _{0}^{1/\kt^{2}} 
          dr_{-} f(r_{-},r_{+}) 
          \left| \frac{\sin \kt r_{-} m}{\kt r_{-}m} \right| 
       +  \int _{1/\kt^{2}}^{\kt^{1 - \epsilon}} dr_{-} f(r_{-},r_{+}) 
          \left| \frac{\sin \kt r_{-} m}{\kt r_{-}} 
          \right|\right.\;\;\nonumber\\
      &+& \left. \int _{\kt^{1 - \epsilon}}^{\infty} dr_{-} f(r_{-},r_{+}) 
          \left| \frac{\sin \kt r_{-} m}{\kt r_{-}} \right| \right]\;\;.
\eea
The last integral doesn't contribute since, for $\epsilon < 1$ and 
large enough $\kt$, $f(r_{-}, r_{+}) = 0$ because $f(r_{-}, r_{+})$ 
is of compact support. The term $\left| \frac{\sin \kt r_{-} m}
{\kt r_{-}m} \right|$ in the first integral can also be bounded because 
it's maximum is 1. Then,
\bea
I_{I} &<&    \frac{C}{\kt^{2}}m \int _{r_{+} < B} dr_{+} 
             \int_{0}^{1/\kt^{2}} dr_{-} f_{\text{max}} 
           + \frac{C}{\kt^{3}} \int _{r_{+} < B} dr_{+} 
             \int_{1/\kt^{2}}^{\kt^{1 - \epsilon}} dr_{-} 
             \frac{f_{\text{max}}}{r_{-}} \;\;\nonumber \\
      &\leq& \frac{D}{\kt^{4}} + \frac{E \ln \kt}{\kt^{3}} \;\;,
\eea
where $f_{\text{max}} = \text{max} f (r_{-}, r_{+})$, and $f = 0$ 
for $r_{+} \geq B$. $D, E$ are appropriate constants. By redefining 
$r_{-}$ to $\bar{r}_{-} := -r_{-}$ we get the same expression for 
$I_{II}$, and therefore the same estimate. This implies that, 
\be
\int_{0}^{m} d\alpha |\beta_{-}|^2
< M\frac{\ln \kt}{\kt^{3}}\;\;,
\ee
for an appropriate M, which again leads to a convergent contribution. 
Since all the terms in Eq.(\ref{betasq}) are finally shown to give 
convergent contributions, the integral in Eq.(\ref{hilbert}) is convergent.

\vspace{0.5in}
\noindent
[Case B.2]  $\kt,\ki \rightarrow \infty;\;\;  \ki^{1/3} \leq 
\kt \leq \ki;\;\;\ki a, \kt a  \gg 1$.\\

Using methods similar to those in [Case B.1] it can be checked that, 
once again, the only potentially divergent contribution is from 
$|\beta_{-}|^2$ with $\beta_{-}$ defined as before, by Eq.(\ref{beta5}). 

To estimate this contribution, we again introduce an auxiliary variable 
$ \alpha := \frac{\ki - \kt}{\kt}\; > 0$ (Note that $\alpha$ in the 
[Case B.2] is different from the $\alpha$ in the [Case B.1] even 
though we use the same symbol.). First, consider the range 
$0 \leq \alpha \leq m$ ($m$ has already been defined, see [Case B.1]).  
Then, as in the [case B.1] we get, for a suitable constant $\tilde{C}$, 
$\tilde{C} > 0$,
\be
\left| \beta_{-} \right| ^{2} \leq
\frac{\tilde{C}}{\kt ^{2}} \int_{a}^{b} \int_{a}^{b} dr_{1} dr_{2}
\frac{T^{\prime}(r_{1})T^{\prime}(r_{2})}{r_{1}r_{2}} e^{i\kt 
\left[ T(r_{1}) - T(r_{2}) \right]}
\left[ \cos \alpha \kt (r_{1} + r_{2})
+ \cos \alpha \kt (r_{1} - r_{2}) \right]\;\;.
\ee
which is exactly the same form as in Eq.(\ref{longeq}). The rest follows 
exactly as in [Case B.1] and the contribution to (\ref{hilbert}) 
from $|\beta_{-}|^{2}$ for $0 \leq \alpha \leq m$ is convergent.

The only remaining case is when $\alpha > m$. We define another 
auxiliary variable $\bar{\alpha} := \frac{\ki - \kt}{\ki} 
= \frac{\kt}{\ki}\alpha$. In the range of $\ki, \kt$ considered 
with $\alpha > m$, $\frac{m}{m+1} < \bar{\alpha} <1$. This implies 
that $0 < 1-\bar{\alpha} < (m+1)^{-1}$. In terms of $\bar{\alpha}$ we have,
\be
\beta_{-} = -\frac{i}{8\pi (\ki\kt)^{1/2}} \int_{a}^{b} dr 
\frac{T^{\prime}}{r}\left[e^{i\ki \left[(1 - \bar{\alpha})T(r) 
+ \bar{\alpha}r \right]} + e^{i\ki \left[(1 - \bar{\alpha})T(r) 
- \bar{\alpha}r \right]}\right]\;\;,
\ee
where $T^{\prime}/r$ is a smooth function of compact support. 
Let $f_{\pm} = (1 - \bar{\alpha})T \pm \bar{\alpha}r$. 
$f_{\pm}^{\prime} \neq 0$ because 
$\left| (1 - \bar{\alpha})T^{\prime} \right| <  
(1 - \bar{\alpha})m < \frac{m}{m + 1} < 
\bar{\alpha}$. Therefore, both terms in the integral fall off 
faster than any power of $1/\ki$ using Lemma 2 $\bf{L. 2. a}$. 
Hence, this contribution from $\beta_{-}$ to (\ref{hilbert}) 
is convergent. This concludes our analysis of [Case B.2].

Since the integral in Eq.(\ref{hilbert}) with $\beta$ given in  
Eq.(\ref{btt}) converges in all domains of integration, we  
conclude that $\beta$ for the half-parametrized case is Hilbert-Schmidt.


\section{Action of Spatial Diffeomorphisms}
\label{sec:spatial}

As discussed in the Introduction, evolution from $\Sigma_{0}$ (with 
radial coordinate `$R$') to an arbitrary axisymmetric slice $\Sigma$ 
(with radial coordinate `$r$') is a composition of `half parametrized' 
evolution followed by the action of a spatial diffeomorphism 
on the final slice. Since `half parametrized' evolution is unitary 
(see section III), evolution from $\Sigma_{0}$ to $\Sigma$ is unitary 
iff the action of spatial diffeomorphisms on $\Sigma$ is unitary. 
As we show in this section, however, the action of a generic 
spatial diffeomorphism on an arbitrary Cauchy slice is not of 
Hilbert-Schmidt type. Therefore, even though `half parametrized' 
evolution is of Hilbert-Schmidt type, generic evolution between 
$\Sigma_{0}$ and $\Sigma$ is not.  

We proceed as follows. First, we show that the expression for 
$\beta(\ki, \kt)$ for a diffeomorphism on $\Sigma$ is identical 
to that for $\beta(\ki, \kt)$ for a corresponding diffeomorphism 
on $\Sigma_{0}$. Next, we show that the explicit expression for 
$\beta(\ki, \kt)$ for generic diffeomorphisms of $\Sigma_{0}$ is 
not Hilbert-Schmidt. 
 
Consider a slice $\Sigma$ with coordinate system $R$ 
(see the remark after Eq.(\ref{nondeg2})). Let `$d$' be 
a diffeomorphism from $\Sigma$ to itself which maps the point 
labelled by $R$ to the point labelled by $r(R)$. Explicitly, 
a point with coordinate $R = R_{0}$ is mapped to a point with 
coordinate $R = r(R_{0})$. The action of `$d$' on Cauchy 
data $(\psi, \pi)$ on $\Sigma$ is a linear canonical transformation 
which can be written as  
\bea
(\psi)_{d}(r) &=& \inty d\tilde{r}  g_{d}(\tilde{r}, r)
                           \psi(\tilde{r})\;\;,\label{diffeo2a}\\ 
(\pi)_{d}(r)  &=& \inty d\tilde{r}  h_{d}(\tilde{r}, r)
                           \pi(\tilde{r})\;\;.\label{diffeo2b}
\eea
Since `$d$' drags the fields at $R$ to $r(R)$, we have that 
$g_{d}(\tilde{r}, r) = h_{d}(\tilde{r}, r) =\delta(\tilde{r},R(r))$. 

Next, consider the initial slice $\Sigma_{0}$ also with coordinate 
system $R$. Let `$d_{0}$' be the diffeomorphism from $\Sigma_{0}$ 
to itself which maps the point labelled by $R$ to the point 
labelled by $r(R)$. Clearly, the action of `$d_{0}$' on data 
$(\psi_{0}, \pi_{0})$ is
\bea
(\psi_{0})_{d_{0}}(r) &=& \inty d\tilde{r}  g_{d_{0}}(\tilde{r}, r)
                           \psi_{0}(\tilde{r})\;\;,\label{diffeoa}\\ 
(\pi_{0})_{d_{0}}(r)  &=& \inty d\tilde{r}  h_{d_{0}}(\tilde{r}, r)
                           \pi_{0}(\tilde{r})\;\;\label{diffeob}
\eea
with $g_{d_{0}}(\tilde{r}, r) = h_{d_{0}}(\tilde{r}, r) 
=\delta(\tilde{r},R(r))$. Thus, for every `$d$' on 
$\Sigma$, there exists `$d_{0}$' on $\Sigma_{0}$ (and vice versa) 
with the same (slice independent) integral kernels. We shall now 
show that, for the Bogolubov transformation of interest, 
$\beta(\ki, \kt)$ for `$d_{0}$' is the same as $\beta(\ki, \kt)$ 
for `$d$'.

First, consider the calculation of $\beta(\ki, \kt)$ corresponding to 
`$d_{0}$'. We compute $\beta(\ki, \kt)$ along the lines described 
in section II. Denote the image of the mode coefficient $a(k)$ 
under `$d_{0}$' as $a_{d_{0}}(k)$. Then, we have that,
\bea
a_{d_{0}}(k) &=& \frac{1}{\sqrt{2}}\inty dr \left[ i \; \jokr 
                        (\pi_{0})_{d_{0}}(r) - r \; \jikr \; 
                        (\psi_{0})_{d_{0}}^{\prime}(r)\; \right]\;\nonumber\\
                    &=& \frac{1}{\sqrt{2}}\inty dr \inty d\tilde{r}
                        \left[ i \; \jokr h_{d_{0}}(\tilde{r}, r)
                        \pi_{0}(\tilde{r}) - r \; \jikr \; 
                        g_{d_{0}}^{\prime}(\tilde{r}, r)
                        \psi_{0}(\tilde{r})\; \right]\;\;.
\label{ckd0}
\eea
Here, we used Eqs.(\ref{diffeoa}) and (\ref{diffeob}) in the second line. 
Finally, we express $(\psi_{0}, \pi_{0})$ in the second line in terms of 
their mode decomposition,  
\bea
\psi_{0}(\tilde{r}) &=& \inty dk \left[ a(k)u(\tilde{r}) 
                     + a^{*}(k)u^{*}(\tilde{r})\right]\label{mode1}\\
\pi_{0} (\tilde{r}) &=& \inty dk \left[ a(k)v(\tilde{r}) 
                     + a^{*}(k)v^{*}(\tilde{r})\right]\label{mode2}\;\;,
\eea
where $u(\tilde{r}) = 1/\sqrt{2} J_{0}(k\tilde{r}) = u^{*}(\tilde{r})$, 
and $v(\tilde{r}) = -i/\sqrt{2} J_{0}(k\tilde{r}) = -v^{*}(\tilde{r})$ 
(see Eq.(\ref{zero})). This allow us to read $\beta$ from
\be
a_{d_{0}}(\ki) = \inty d\kt \left[ \alpha(\ki,\kt)\;a(\kt) 
+ \beta(\ki,\kt)\;a^{*}(\kt) \right] \;\;.
\ee
The computation of $\beta$ for an arbitrary slice $\Sigma$ is an 
exactly analogous process. The action of `$d$' on data $(\psi, \pi)$ 
is given by (\ref{diffeo2a}) and (\ref{diffeo2b}). 
Here, $(\psi, \pi)$ is the image of initial data on $\Sigma_{0}$ 
under half parametrized evolution. In section III we showed that 
the Bogolubov transformation for half parametrized evolution 
between $a(k)$ and $a_{\Sigma}(k)$ was Hilbert-Schimidt. 
It remains to evaluate the action of the canonical transformation 
(\ref{diffeo2a}) and (\ref{diffeo2b}) on these $a_{\Sigma}(k)$. 
From section II, $a_{\Sigma}(k)$ and $(\psi, \pi)$ are related by
\be
a_{\Sigma}(k) = \frac{1}{\sqrt{2}}\inty dr \left[ i \; \jokr 
                        \pi(r) - r \; \jikr \; 
                        \psi^{\prime}(r)\; \right]\;\;, 
\label{asigma2}
\ee
and
\bea
\psi(\tilde{r}) &=& \inty dk \left[ a_{\Sigma}(k)u(\tilde{r}) 
                     + a_{\Sigma}^{*}(k)u^{*}(\tilde{r})\right]
                    \label{mode3}\\
\pi(\tilde{r}) &=& \inty dk \left[ a_{\Sigma}(k)v(\tilde{r}) 
                     + a_{\Sigma}^{*}(k)v^{*}(\tilde{r})\right]
                   \label{mode4}\;\;.  
\eea
From Eqs.(\ref{diffeo2a}),(\ref{diffeo2b}) and (\ref{asigma2}), the 
image of $a_{\Sigma}(k)$ under `$d$' is  
\bea
(a_{\Sigma})_{d}(k) &=& \frac{1}{\sqrt{2}}\inty dr \left[ i \; \jokr 
                        (\pi)_{d}(r) - r \; \jikr \; 
                        (\psi)_{d}^{\prime}(r)\; \right]\;\nonumber\\
                    &=& \frac{1}{\sqrt{2}}\inty dr \inty d\tilde{r}
                        \left[ i \; \jokr h_{d}(\tilde{r}, r)
                        \pi(\tilde{r}) - r \; \jikr \; 
                        g_{d}^{\prime}(\tilde{r}, r)
                        \psi(\tilde{r})\; \right]\;\;.
\label{ckd}
\eea
Substituting Eqs.(\ref{mode3}) and (\ref{mode4}) into Eq.(\ref{ckd}), 
we can read off $\beta$ from
\be
(a_{\Sigma})_{d}(\ki) = \inty d\kt \left[ \alpha(\ki,\kt)\;
a_{\Sigma}(\kt) 
+ \beta(\ki,\kt)\;a_{\Sigma}^{*}(\kt) \right] \;\;.
\ee
Since $h_{d} = h_{d_{0}}$, $g_{d} = g_{d_{0}}$, $\beta$ is identical 
to the expression
obtained when $\Sigma = \Sigma_{0}$. Therefore, without loss of 
generality, we shall use $\beta$ corresponding to the action 
of `$d_{0}$' on the initial slice $\Sigma_{0}$.

$\beta(\ki, \kt)$ for the initial slice $\Sigma_{0}$ can be easily 
computed by setting $T = 0$, and $T^{\prime} = 0$ in Eq.(\ref{betagen}) so that
\be
\beta (\ki,\kt) = \frac{\kt}{2}\inty \; dR 
\left[ r\jikir\jiktR - R\jokir\joktR \right] \;\;.
\label{beta2}
\ee
Note that $r\jikir - R\jikiR$ and $\jokir -\jokiR$ have support 
only in $[c,d]$ since $R=r$ outside $[c,d]$, Eq.(\ref{bcr}). 
This together with Eq.(\ref{orthogonal}) implies that, for a smooth function 
$g(R)$ with compact support in 
$[\bar{c},\bar{d}]$, where $\bar{c} < c < d < \bar{d}$, 
such that $g(R)=1$ for $R \in [c,d]$, we can write $\beta$ as 
\bea
\beta (\ki,\kt) 
&=& \frac{\kt}{2}\int_{\bar{c}}^{\bar{d}} \; dR g(R) 
    \left[ r\jikir\jiktR - R\jokir\joktR \right] \label{betafin1}\\
&-& \frac{\kt}{2}\int_{\bar{c}}^{\bar{d}} \; dR g(R) R
    \left[ \jikiR\jiktR - \jokiR\joktR \right] \label{betafin2}\\
&:=&\beta_{[r,R]} (\ki,\kt) - \beta_{[R,R]} (\ki,\kt)\;\;.\label{betafin3}
\eea

To show that the integral in Eq.(\ref{hilbert}) with $\beta$ given 
in (\ref{betafin1}) to (\ref{betafin3}) diverges, it is sufficient 
to show that the integral diverges over any subset of the domain of 
integration because the integrand in Eq.(\ref{hilbert}) is always 
positive. We choose this subset as follows. We change variables 
from $(\ki, \kt)$ to $(\alpha:=\ki/\kt, \kt)$. Let 
$\alpha_{0} \in \bf{R}$ be such that $\alpha_{0} > 0,\; \alpha_{0} \neq 1$. 
We shall restrict attention to the subdomain of integration 
defined by $\alpha_{0} - \delta < \alpha < \alpha_{0} + \delta$ 
and $\kto<\kt < 2\kto$, $\kto \rightarrow \infty$. 
Here $\delta$ is chosen such that $\delta \ll \alpha_{0}$ so that, 
in the range considered, $\alpha > 0$ and $\alpha \neq 1$. 

Similar to the previous section, we use the asymptotic expansion of 
Bessel function for large arguments, Appendix B, to estimate $\beta$. 
We get,
\bea
\beta_{[r,R]} 
&=&\frac{1}{4\pi\alpha^{1/2}} \int_{\bar{c}}^{\bar{d}} \;dR g(R) 
      \left[ \frac{r+R}{\sqrt{rR}}i -\frac{(3r-R)(R+\alpha r)}
      {8\alpha\sqrt{(rR)^{3}}}\frac{1}{\kt} \right.\nonumber\\
&+& \left.\frac{1}{128}\left(
      \frac{(15r - 9R)(R^2 + \alpha^2r^2)}{\alpha^2\sqrt{(rR)^5}}-
      \frac{(18r+2R)}{\alpha\sqrt{(rR)^3}}\right)i\frac{1}{\kt^2}
      \right]e^{i\kt (R + \alpha r)} 
      + c.c. \label{beta3a} \\ 
&+&   \frac{1}{4\pi\alpha^{1/2}} \int_{\bar{c}}^{\bar{d}} \;dR g(R) 
      \left[ \frac{r-R}{\sqrt{rR}} -\frac{(3r+R)(R-\alpha r)}
      {8\alpha\sqrt{(rR)^{3}}}i\frac{1}{\kt}\right.\nonumber\\
&+&   \left. \frac{1}{128}\left(
      \frac{(15r + 9R)(R^2 + \alpha^2r^2)}{\alpha^2\sqrt{(rR)^5}}+
      \frac{(18r-2R)}{\alpha\sqrt{(rR)^3}}\right)\frac{1}{\kt^2}\right] 
      e^{i\kt (R - \alpha r)}
      + c.c. \label{beta3b} \\
&+&   O\left(\frac{1}{\kt^{3}}\right)\\
&:=&  \beta^{+}_{[r,R]} + \beta^{-}_{[r,R]}\label{beta3c} 
      + O\left(\frac{1}{\kt^{3}}\right)\;\;,
\eea	
where `$c.c$' refers to `complex conjugate', $\beta^{+}_{[r,R]}$ is 
(\ref{beta3a}) and $\beta^{-}_{[r,R]}$ is (\ref{beta3b}). 
$\beta_{[R,R]}$ in Eq.(\ref{betafin3}) can also be written 
down in a similar form by replacing $r$ with $R$ in (\ref{beta3a}) and 
(\ref{beta3b}). Similar to Eq.(\ref{beta3c}) we can also write 
$\beta_{[R,R]}$ as a sum of $\beta^{+}_{[R,R]}$, $\beta^{-}_{[R,R]}$ 
and a term of $O(1/\kt^{3})$. Then, $\beta$ is,
\be
\beta = \beta^{+}_{[r,R]} + \beta^{-}_{[r,R]} + \beta^{+}_{[R,R]} 
      + \beta^{-}_{[R,R]} + O\left( \frac{1}{\kt^3}\right)\;\;.
\label{KKL}
\ee

Now, we show that $\beta^{+}_{[R,R]}$, $\beta^{-}_{[R,R]}$, and 
$\beta^{+}_{[r,R]}$ are of rapid decrease in $\kt$. 
Consider $\beta^{+}_{[r,R]}$ first. The integrals in $\beta^{+}_{[r,R]}$ 
are of the form in Lemma 2, $\bf{L.2.a}$, because $R^{\prime} > 0$ 
and $\alpha >0$ in the domain of interest implies that
\be
\frac{d}{dR}(R+\alpha r) = 1+\alpha\frac{dr}{dR} \neq 0 \;\;.
\label{momo}
\ee
A similar argument holds for $\beta^{+}_{[R,R]}$. The contribution 
from $\beta^{-}_{[R,R]}$ can also be estimated using 
Lemma 2, $\bf{L.2.a}$. Explicitly, $\beta^{-}_{[R,R]}$ is,
\be
\frac{1}{4\pi\alpha^{1/2}} \int_{\bar{c}}^{\bar{d}} \;dR g(R) 
      \left[ \frac{(\alpha-1)}
      {2\alpha R}i\frac{1}{\kt}
+   \frac{1}{16R^2}
      \frac{3\alpha^2 + 2\alpha +3}{\alpha^2}\frac{1}{\kt^2}
      \right] 
      e^{i\kt (1- \alpha)R}
      + c.c. \label{betaRR}
\ee
Since $\alpha \neq 1$, we have that 
\be
\frac{d}{dR}(1 - \alpha)R \neq 0 \;\;.
\ee
Therefore, $\beta^{-}_{[R,R]}$ is of the form in Lemma 2 $\bf{L.2.a}$ 
in the domain of interest. 

Since $\beta^{+}_{[R,R]}$, $\beta^{-}_{[R,R]}$ and $\beta^{+}_{[r,R]}$ 
are all of the form in Lemma 2 $\bf{L.2.a}$, they are functions of 
rapid decrease in $\kt$. Then, from (\ref{KKL}), we get, 
\be
\beta = \beta^{-}_{[r,R]} + O\left(\frac{1}{\kt^{3}}\right)\;\;.
\ee
A rough estimate of $\beta^{-}_{[r,R]}$ from (\ref{beta3b}) 
gives $\beta^{-}_{[r,R]} = O(1)$. This implies that,
\be
|\beta|^{2} = |\beta^{-}_{[r,R]}|^{2} + O\left(\frac{1}{\kt^{3}}\right)\;\;.
\label{KKLL}
\ee
A term of $O(1/\kt^{3})$ gives a convergent contribution to the 
integral in Eq.(\ref{hilbert}), and it only remains to estimate 
$\beta^{-}_{[r,R]}$ with more care.

As in \cite{charlie2}, we show (for a large class of embeddings $R(r)$) 
that a judicious use of the stationary phase approximation to 
estimate $\beta^{-}_{[r,R]}$ obtains a divergent contribution to 
(\ref{hilbert}). Specifically, we demand that the embedding (and our 
choice of $\alpha_{0}$) be such that the following ``genericity'' 
conditions hold.\\
\vspace{0.05in}

\noindent
(a) There exists a number $\alpha_{0} \neq 1$ with the property that 
$R^{\prime} = \alpha_{0}$ at only finitely many points 
$r_{I},\;I=1,...,N$ in $[c,d]$.\\
(b) $R^{\prime\prime} \neq 0$ at $r_{I}$.\\
(c) At least one of $\tilde{\tilde{f}}_{I}$, 
$I=1, ..., \tilde{\tilde{N}}$, (see(\ref{betacKK}) below) is non-zero.\\
\vspace{0.05in}

\noindent
Condition (a) implies that there exists a fixed finite 
number of critical points for the exponent in (\ref{beta3b}) at 
$\alpha = \alpha_{0}$. Note that the condition for critical points 
is $R^{\prime} = \alpha$. This condition defines a map `$\lambda$' 
from $r \in \Sigma$ to $\alpha \in (\alpha_{0} - \delta,\; 
\alpha_{0} + \delta)$. From this point of view, (b) states that the 
differential of this map is non-degenerate at $r=r_{I}$ 
(with $\lambda(r_{I}) = \alpha_{0}$). An application of the inverse mapping 
theorem to $\lambda$ implies that the number of critical points $N$ 
does not change as $\alpha$ varies in a sufficiently small neighborhood 
of $\alpha_{0}$ (i.e. if $\delta$ is chosen small enough) and that the 
location of the critical points varies continuously with $\alpha$ in this 
neighborhood. 

After these preliminaries, we proceed with our estimate 
for $\beta^{-}_{[r,R]}$. The leading order term in (\ref{beta3b}) is,
\be
 \frac{1}{4\pi\alpha^{1/2}} \int_{\bar{c}}^{\bar{d}} \;dR \; g(R) 
    \left( \frac{r-R}{\sqrt{rR}}
    \right)e^{-i\kt (\alpha r-R)} 
    + c.c. \;\;.
\label{leading}
\ee
We apply the stationary phase approximation to estimate (\ref{leading}). 
As discussed above, there are $N$ critical points $r_{I}$ where 
$R^{\prime}=\alpha$ and we estimate (\ref{leading}) as  
\be
\frac{1}{4\pi\alpha^{1/2}}\left(\frac{2\pi}{\kt}\right)^{1/2} 
\sum_{I = 1}^{N} f_{I}  e^{-i\kt G_{I}} + c.c.
+ O\left(\frac{1}{\kt^{3/2}}\right)\;\;,
\label{betastationary}
\ee
where
\bea
f_{I} &:=& \left| \det R^{\prime\prime}(r_{I})\right|^{-1/2} 
           \left( \frac{r_{I}-R(r_{I})}{\sqrt{r_{I}R(r_{I})}}
           \right)e^{i\pi\text{sign} 
           R^{\prime\prime}/4}\label{genericf}\\
G_{I} &:=& \alpha r_{I} - R(r_{I})\label{genericg}\;\;. 
\eea
Note that $r_{I}$ depend on $\alpha$, i.e. $r_{I} = r_{I}(\alpha)$. 

The stationary phase approximation applied to the subleading order 
terms in (\ref{beta3b}) yields an estimate of $O(1/\kt^{3/2})$, 
which is the same order as the ``error term'' in Eq.(\ref{betastationary}). 
Then, we have
\be
\beta^{-}_{[r,R]}= 
\frac{1}{4\pi\alpha^{1/2}}\left(\frac{2\pi}{\kt}\right)^{1/2} 
\sum_{I = 1}^{2N} \tilde{f}_{I} e^{-i\kt \tilde{G}_{I}}
+ O\left(\frac{1}{\kt^{3/2}}\right)\;\;,
\label{betastationary2}
\ee
where, for $I = 1,2, ..., N$, $\tilde{f}_{I} = f_{I}$, $\tilde{G}_{I} 
= G_{I}$, $\tilde{f}_{I+N} = f_{I}^{*}$, $\tilde{G}_{I+N} = - G_{I}$, with 
$f_{1},...,f_{N}$ and $G_{1},...,G_{N}$ defined in Eqs.(\ref{genericf}) 
and (\ref{genericg}). It is possible that $\tilde{G}_{I} = \tilde{G}_{J}$ 
for some $I \neq J$, $I, J = 1,2, ..., 2N$. By suitably redefining the 
$\tilde{f}_{I}$, with the redefined $\tilde{f}_{I}$ denoted by 
$\tilde{\tilde{f}}_{I}$, the sum $\sum_{I = 1}^{2N} \tilde{f}_{I} 
e^{-i\kt \tilde{G}_{I}}$ can be rewritten as a sum over a finite number 
$\tilde{\tilde{N}} < 2N$ of terms, each of the form $\tilde{\tilde{f}}_{I} 
e^{-i\kt \tilde{\tilde{G}}_{I}}$, $I = 1,2, ..., \tilde{\tilde{N}}$, 
$\tilde{\tilde{G}}_{I} \neq \tilde{\tilde{G}}_{J}$ for $I \neq J$. 
Then, we get,
\be
| \beta^{-}_{[r,R]} |^{2} 
> \frac{1}{8\pi\kt\alpha} \left( \sum_{I}^{\tilde{\tilde{N}}} 
  | \tilde{\tilde{f}}_{I} |^{2}
    + \sum_{I \neq J}^{\tilde{\tilde{N}}} \tilde{\tilde{f}}_{I}  
    \tilde{\tilde{f}}^{*}_{J}e^{-i\kt 
    (\tilde{\tilde{G}}_{I} - \tilde{\tilde{G}}_{n})} \right) 
+ O\left(\frac{1}{\kt^{2}}\right)\;\;.
\label{betacKK}
\ee
For generic embeddings, at least one of the $\tilde{\tilde{f}}_{I} 
\neq 0$ and as mentioned in (c) above, we restrict our attention 
to such embeddings.

Now, we are ready to estimate the integral in Eq.(\ref{hilbert}) 
in the domain of interest. Using Eqs.(\ref{KKLL}) and (\ref{betacKK}) 
the integral is,
\bea
\int_{\kto}^{2\kto} &\kt& d\kt 
\int_{\alpha_{0}-\delta}^{\alpha_{0}+\delta}
d\alpha |\beta(\ki,\kt)|^{2}\\
&>&
\frac{1}{2}\int_{\kto}^{2\kto} \kt d\kt 
\int_{\alpha_{0}-\delta}^{\alpha_{0}+\delta}
d\alpha| \beta^{-}_{[r,R]}|^{2} \label{fin}\\
&>& 
\frac{1}{16\pi}\int_{\kto}^{2\kto} d\kt 
\int_{\alpha_{0}-\delta}^{\alpha_{0}+\delta}
\frac{d\alpha}{\alpha}\sum_{I} | \tilde{\tilde{f}}_{I} |^{2}\label{fina}\\ 
&-&
\left|\frac{1}{16\pi}\int_{\kto}^{2\kto} d\kt 
\int_{\alpha_{0}-\delta}^{\alpha_{0}+\delta} \frac{d\alpha}{\alpha}
\sum_{I \neq J}^{\tilde{\tilde{N}}} \tilde{\tilde{f}}_{I}
\tilde{\tilde{f}}^{*}_{J}e^{-i\kt(\tilde{\tilde{G}}_{I} - 
\tilde{\tilde{G}}_{n})}\label{finc}\right|\\ 
&-&
\left|\frac{1}{16\pi}\int_{\kto}^{2\kto} d\kt 
\int_{\alpha_{0}-\delta}^{\alpha_{0}+\delta} \frac{d\alpha}{\alpha}
\frac{C}{\kt}\right| 
    \label{finb}\;\;,
\eea
for some appropriate $C>0$. 
Since at least one $\tilde{\tilde{f}}_{I} \neq 0$, (\ref{fina}) 
is equal to $A\kto$ for some suitably defined positive constant $A$. 
(Note that $A$ depends on $\alpha_{0}$ and $\delta$ but is independent of 
$\kto$). Eq.(\ref{finc}) is estimated as of $O(1)$ (in $\kto$) 
by integrating over $\kt$ to get
\be
\left.\int_{\alpha_{0}-\delta}^{\alpha_{0}+\delta} \frac{d\alpha}{\alpha}
\sum_{I \neq J}^{\tilde{\tilde{N}}} 
\frac{\tilde{\tilde{f}}_{I}\tilde{\tilde{f}}^{*}_{J}}
{(\tilde{\tilde{G}}_{I} - \tilde{\tilde{G}}_{J})}
ie^{-i\kt(\tilde{\tilde{G}}_{I} - \tilde{\tilde{G}}_{J})}
\right|_{\kto}^{2\kto}
= O(1)\;\;.
\ee
Similarly an integration over $\kt$ show that (\ref{finb}) is also 
of $O(1)$. All this implies that, for large enough $\kto$
\be
\int_{\kto}^{2\kto} \kt d\kt 
\int_{\alpha_{0}-\delta}^{\alpha_{0}+\delta}
d\alpha |\beta(\ki,\kt)|^{2} > \frac{A}{2}\kto\;\;,
\ee
which diverges as $\kto\rightarrow \infty$. Thus, we have shown 
that the contribution to (\ref{hilbert}) from the domain of interest is 
(linearly) divergent. Therefore, the action of a generic spatial 
diffeomorphism is not of Hilbert-Schmidt type.


\section{ Nonexistence of Infinitesimal generators of Spatial Diffeomorphism}
\label{sec:inf}

After our analysis of finite spatial diffeomorphisms 
in the previous section, we now turn to a study of 
infinitesimal diffeomorphisms. Specifically, we are 
interested in investigating if the quantum operators 
corresponding to generators of finite spatial diffeomorphisms 
are densely defined on the dense set consisting of 
the Fock vacuum and suitably defined N-particle states. 
Preliminary computations along the lines of \cite{me} 
suggest that the generator of a spatial diffeomorphism 
is well-defined on this dense domain iff its action on 
the Fock vacuum is well-defined. Hence, we investigate 
if the action of such a generator is well-defined on 
the Fock vacuum. 

Note that this section constitutes a digression from 
the main theme of this paper in that there is no direct 
relation between the existence (or lack thereof) of the 
action of such a generator on the Fock vacuum and the 
lack of unitary implementability of generic finite spatial 
diffeomorphisms. In particular, even if the generator 
turned out to be a densely-defined self-adjoint operator, 
it would not necessarily imply the existence of a unitary 
operator corresponding to the finite diffeomorphisms it 
(putatively) generates because there could be operator 
domain problems in defining (``path-ordered") exponentiation. 
Conversely, had finite diffeomorphisms been unitarily 
implementable, this would not necessarily imply that 
the infinitesimal generators were defined as self-adjoint 
operators or, even if they were, that the vacuum was in 
their dense domain.

Hence, our purpose is merely to initiate an exploration 
of properties of the generator of spatial diffeomorphisms 
in the hope that such an exploration may be useful for 
further work in the field. The result of our computations 
is that operators corresponding to generic generators of 
spatial diffeomorphisms \footnote{The notion of genericity 
used here is defined later in this section and in Appendix 
\ref{sec:genericity}. We do not know how it relates to the 
genericity conditions of section IV.} of the flat initial 
slice $\Sigma_0$ do not have a well-defined action on the 
Fock vacuum. The generator of spatial 
diffeomorphisms of a slice $\Sigma$ in classical theory is 
$\int_{\Sigma} dr f(r)\pi\psi^{\prime}$ where $f(r)$ is the radial shift 
vector. When $\Sigma = \Sigma_{0}$, a straightforward computation shows 
that the action of the corresponding normal ordered quantum operator on 
the Fock vacuum state is,
\be
\| \int_{0}^{\infty} dr\; rf(r)\;:
\widehat{\dot{\psi}(r,0)}\widehat{\psi'(r,0)}: 
| 0 \rangle \|^2 = \frac{1}{8}\inty d\ki \inty d\kt 
\left| F(\ki\kt) \right|^2 \,\,,
\label{sinfty}
\ee
where 
\be
F(\ki,\kt) = \ki\kt\inty dr\;rf(r)\left[\jokir\jiktr + \jikir\joktr \right].
\label{F}
\ee
Here the radial shift, $f(r)$, is assumed to be of compact 
support away from $r=0$. We use Eq.(\ref{psi}) and the 
canonical commutation relations between the $\hat{a}(k)$ 
and $\hat{a}^{\dagger} (k)$ in the left hand side of 
Eq.(\ref{sinfty}) to get Eq.(\ref{F}). If the integral on 
the right hand side of Eq.(\ref{sinfty}) diverges, the action 
of the generator is not well-defined. In the rest of the section 
we show that this is indeed the case. 

To show that the integral in Eq.(\ref{sinfty}) diverges, 
we only need to show that it does so for a sub-domain of 
integration because the integrand is always positive. 
We restrict attention to the sub-domain $\kt > \ki$, 
$\kt - \ki = a$ with $a$ of $O(1)$ and $\ki, \kt \rightarrow \infty$. 
Using the Hankel transformation, $G(k)$, of $rf(r)$ with respect 
to $\jokr$ (see Eq.(\ref{g})) and identities in Appendix 
$\ref{sec:appB}$, we get 
\be
F(\ki, \kt) = \ki\kt \int_{\kt-\ki}^{\kt+\ki} dk\;kG(k) 
\left[ \frac{\theta (\ki, k ; \kt)}
{\pi\kt} + \frac{\theta (\kt, k ; \ki)}{\pi\ki} \right] 
+ \ki\int_{0}^{\kt - \ki} dk\;kG(k)\;,
\label{hankelF}
\ee
where $\theta (\ki, k ; \kt)$ is the angle between $\ki$ and $k$ 
in the triangle formed by $\ki, k$ and $\kt$, and similarly for 
$\theta (\kt, k ; \ki)$\footnote{The notation used here is following. 
$\theta (k_{a}, k_{b}; k_{c})$ is defined as the angle between 
$k_{a}, k_{b}$ in the triangle formed by $k_{a}, k_{b}$ and $k_{c}$.}. 
By adding and subtracting terms, and using $\theta (\kt, k ; \ki) + 
\theta (\ki, k ; \kt) + \theta (\kt, \ki ; k) = \pi$, we get
\bea
F(\ki,\kt) 
&=& \ki \int_{0}^{\ki + \kt} dk\; kG(k)\label{I1}\\
&-& \ki\int_{\kt-\ki}^{\kt+\ki} dk\;kG(k) 
    \frac{\theta (\kt, \ki ; k)}{\pi}\label{I2}\\ 
&+& \int_{\kt-\ki}^{\kt+\ki} dk\;kG(k)(\kt - \ki)
    \frac{\theta (\kt, k ; \ki)}{\pi}\label{I3}\;.
\eea

Before we jump into the rather lengthy estimate of each term, 
it would be helpful to state the main results here. 
It turns out that, for sufficiently generic embeddings, 
the leading contribution to $F(\ki, \kt)$ comes from (\ref{I2}) 
in the domain we are interested in. Using the estimate 
of (\ref{I2}) we show that the right hand side of Eq.(\ref{sinfty}) 
diverges.  Contributions from both (\ref{I1}) and (\ref{I3}) 
turn out to be negligible compared to the one from (\ref{I2}).

First, we estimate the integral in Eq.(\ref{I1}). It can be written as 
\be
\ki\int_{0}^{\ki + \kt} dk\; kG(k) = \ki\int_{0}^{\infty} dk\; kG(k)
                                   - \ki\int_{\ki + \kt}^{\infty} dk\; kG(k)\;.
\label{I1a}
\ee
The first integral on the right hand side is zero because, 
from Eq.(\ref{hankel}), it is equal to $rf(r)\arrowvert_{r=0}$, and  $f(r)$ 
does not have support at $r=0$\footnote{The Hankel transformation 
with respect to $\jokr$ can be thought of as a two dimensional 
Fourier transformation. Since $f(r)$ is Schwartz, so is its 
Fourier transform, $G(k)$. See Appendix A.1 of \cite{me}.}. 
Moreover, as shown in Appendix A.1 of \cite{me}, $G(k)$ 
falls faster than any power of $1/k$ as $k \rightarrow \infty$. 
Since $k >(\ki + \kt)$, $k\rightarrow\infty$, and the contribution 
from the second integral is negligible.

Next, consider Eq.(\ref{I2}). There are 
two contributions, namely, from $G(k)$ 
and from $\theta(\ki,\kt;k)$. Since $G(k)$ falls off 
rapidly when $k\rightarrow\infty$, and $\theta(\ki,\kt;k)$ 
is bounded, large contributions can arise only when $k$ 
is small. So, we need to estimate $\theta(\ki,\kt;k)$ 
when $k$ is small. 

Since $\ki,\kt$ and $k$ form a triangle, we have in the 
domain of interest that
\bea
k^2 &=& (\ki - \kt)^2 + 4\ki\kt\sin^2\frac{\theta}{2}\nonumber\\
    &=& a^2 + 4\ki\kt\sin^2\frac{\theta}{2}\;.
\label{sin}
\eea
Since $0<\theta<\pi$, $k$ increases when $\theta$ increases, for 
fixed $\ki, \kt$. Therefore, $k$ is small when $\theta$ is small. 
For small $\theta$ we can expand $\sin\frac{\theta}{2}$ in 
Eq.(\ref{sin}) to get
\be
k^2 = a^2 + 4\ki\kt \left[\frac{\theta^2}{4} - \frac{\theta^4}{48}
      + O(\theta^6) \right]\;.
\label{theta}
\ee

Now, to see how small $\theta$ and $k$ should be to have a non-trivial 
contribution, we divide the domain into two regions; 
$\theta \geq (\ki\kt)^{-1/2 +\epsilon}$ and 
$\theta < (\ki\kt)^{-1/2 +\epsilon}$ where $0 < \epsilon \ll 1/2$. 
If $\theta = (\ki\kt)^{-1/2 +\epsilon}$, then $k^2 =
(\ki\kt)^{2\epsilon} + O(1)$. 
Thus, $k\rightarrow\infty$ when $\ki,\kt\rightarrow\infty$. 
Since $k$ increases with $\theta$, the contribution from 
$\theta \geq (\ki\kt)^{-1/2 +\epsilon}$ is negligible due to 
the nice fall off behavior of $G(k)$ at large $k$.

Consider the contribution to Eq.(\ref{I2}) from 
\be
a< k < \sqrt{a^2 + (\ki\kt)^{2\epsilon}}\;.
\label{a}
\ee
This implies that $0 < 4\sin^{2}\theta/2 < (\ki\kt)^{-1 +2\epsilon}$ 
so that the upper bound of $\theta$ in this range exceeds 
$(\ki\kt)^{-1/2 + \epsilon}$ by $O((\ki\kt)^{-3/2 + 3\epsilon})$. 
Since the contribution from the entire range 
$\theta \geq (\ki\kt)^{-1/2 +\epsilon}$ is negligible, it makes 
no difference whether we estimate the contribution from 
$\theta \leq (\ki\kt)^{-1/2 +\epsilon}$ or from Eq.(\ref{a}) and 
we use the latter.

In this range, for large enough $\ki, \kt$, we have from 
Eq.(\ref{theta}) that
\be
\theta = \left(\frac{k^2 - a^2}{\ki\kt}\right)^{1/2} 
         + O\left(\frac{1}{(\ki\kt)^{3/2 - 3\epsilon}}\right)\;.
\label{thetaa}
\ee
Then, Eq.(\ref{I2}) satisfies the bounds
\bea
\left|\ki\int_{\kt-\ki}^{\kt+\ki} dk\;kG(k) 
\frac{\theta (\kt, \ki ; k)}{\pi}\right|
&>& \left| \ki\int_{a}^{\sqrt{a^2 + (\ki\kt)^{2\epsilon}}}
    dk\;kG(k) \frac{\theta (\kt, \ki ; k)}{\pi} \right| \nonumber \\
&-& \left| \ki\int_{\sqrt{a^2 + (\ki\kt)^{2\epsilon}}}^{\ki +\kt}
    dk\;kG(k) \frac{\theta (\kt, \ki ; k)}{\pi} \right| \nonumber \\
&>& \left| \frac{\ki}{\pi}\int_{a}^{\sqrt{a^2 + (\ki\kt)^{2\epsilon}}} 
    dk\;kG(k)\left(\frac{k^2 - a^2}{\ki\kt}\right)^{1/2} \right| \nonumber \\
&-& \left| \frac{\ki}{\pi}\int_{a}^{\sqrt{a^2 + (\ki\kt)^{2\epsilon}}} 
    dk\;kG(k) O\left(\frac{1}
    {(\ki\kt)^{3/2-3\epsilon}}\right)\right|\nonumber \\
&-& \left| \ki\int_{\sqrt{a^2 + (\ki\kt)^{2\epsilon}}}^{\ki +\kt}
    dk\;kG(k) \frac{\theta (\kt, \ki ; k)}{\pi} \right| \nonumber \\
&:=&\left| I_{2} \right| - \left | I_{3} \right| - \left| I_{1} \right| \; ,
\label{I2a}
\eea  
where we have used (\ref{thetaa}).
As discussed above, $\left | I_{1} \right| $ is negligible. 
On the other hand, we have that 
\bea
\left| I_{3} \right| 
&=& \frac{\ki}{\pi(\ki\kt)^{3/2}}\left|
    \int_{a}^{\sqrt{a^2 + (\ki\kt)^{2\epsilon}}}dk\;kG(k) O
    \left[(\ki\kt)^{3\epsilon}\right] \right| \nonumber \\
&<& \frac{C_{1}\ki}{\pi(\ki\kt)^{3/2-3\epsilon}} 
    \int_{a}^{\sqrt{a^2 + (\ki\kt)^{2\epsilon}}}dk\;k\left|G(k)\right|
    \nonumber \\
&<& \frac{C_{1}\ki}{\pi(\ki\kt)^{3/2 - 3\epsilon}}
    \int_{0}^{\infty}dk\;k\left|G(k)\right|\nonumber \\
&=&\frac{\ki}{(\ki\kt)^{3/2 - 3\epsilon}} C_{2} \;\label{D},
\eea
for appropriate constant $C_{1}$ and $C_{2}$. 
Finally $\left| I_{2} \right|$ is,
\bea
\left| I_{2} \right| 
&=& \frac{\ki}{\pi}\left|
    \int_{a}^{\sqrt{a^2 + (\ki\kt)^{2\epsilon}}}dk\;kG(k)
    \left(\frac{k^2 - a^2}{\ki\kt}\right)^{1/2} \right| \nonumber \\ 
&=& \frac{1}{\pi}\sqrt{\frac{\ki}{\kt}} \left| 
    \int_{a}^{\sqrt{a^2 + (\ki\kt)^{2\epsilon}}}dk\;kG(k)\sqrt{k^2 - a^2} 
    \right| \nonumber \\
&:=&\frac{1}{\pi}\sqrt{\frac{\ki}{\kt}} \left| C(a,\ki, \kt) \right| 
    \label{I2b}\;.
\eea

If $C(a,\ki, \kt) \neq 0$ as $\ki, \kt \rightarrow\infty$ it follows that 
$I_{2}$ dominates over $I_{3}$ in this limit. 
The leading contribution to (\ref{I2}) is then (\ref{I2b}). 
We now show that, for sufficiently generic shift vector $f(r)$ 
(see Appendix \ref{sec:genericity} for the definition of genericity), 
this is indeed the case. The idea is to estimate $ C(a,\ki, \kt) 
$ as $a \rightarrow 0$ and $\ki, \kt \rightarrow \infty$. 
Consider the behavior of $C(a,\ki, \kt)$ 
in the region where $a \in \left[\frac{1}{\kt^{\delta}}, 
\frac{2}{\kt^{\delta}}\right],\; 0<\delta \ll 1$ and 
$\ki, \kt \rightarrow \infty$.
Define $\sigma$ as $\sigma := a\kt^{\delta}$. Then,
\be
C(a,\ki, \kt) = \int_{\frac{\sigma}{\kt^{\delta}}}^{(\ki\kt)^{\epsilon}}
                dk\;kG(k)\sqrt{k^2 - (\frac{\sigma}{\kt^{\delta}})^2}
              + \int_{(\ki\kt)^{\epsilon}}^{\sqrt{(\sigma/\kt^{\delta})^2 
              + (\ki\kt)^{2\epsilon}}}
                dk\;kG(k)\sqrt{k^2 - (\frac{\sigma}{\kt^{\delta}})^2}\;\;.
\label{ccc}
\ee
The second integral in Eq.(\ref{ccc}) is negligible since $G(k)$ falls 
faster than any power of $1/k$\cite{me}. The first integral in 
Eq.(\ref{ccc}) can be estimated as,
\bea
       \int_{\frac{\sigma}{\kt^{\delta}}}^{(\ki\kt)^{\epsilon}}
       &dk&\;kG(k)\sqrt{k^2 - (\frac{\sigma}{\kt^{\delta}})^2}
       \nonumber \\
&=&    \int_{\frac{\sigma}{\kt^{\delta}}}^{\frac{\sigma}
       {\kt^{\delta/2}}}
       dk\;kG(k)\sqrt{k^2 - (\frac{\sigma}{\kt^{\delta}})^2}
 +     \int_{\frac{\sigma}{\kt^{\delta/2}}}^{(\ki\kt)^{\epsilon}}
       dk\;kG(k)\sqrt{k^2 - (\frac{\sigma}{\kt^{\delta}})^2}
       \nonumber \\ 
&=&    O(\frac{1}{\kt^{\delta/2}})
+      \int_{\frac{\sigma}{\kt^{\delta/2}}}^{(\ki\kt)^{\epsilon}}
             dk\;k^2G(k)+ 
       \int_{\frac{\sigma}{\kt^{\delta/2}}}^{(\ki\kt)^{\epsilon}}
             dk\;k^2G(k)O(\frac{1}{\kt^{\delta}}) \nonumber\\ 
&=&   O(\frac{1}{\kt^{\delta/2}})
+     \int_{\frac{\sigma}{\kt^{\delta/2}}}^{(\ki\kt)^{\epsilon}}
      dk\;k^2G(k)  + O(\frac{1}{\kt^{\delta}})\nonumber\\ 
&=&    O(\frac{1}{\kt^{\delta/2}})
 +     \int_{\frac{\sigma}{\kt^{\delta/2}}}^{(\ki\kt)^{\epsilon}}
                     dk\;k^2 G(k)\;\;. 
\eea
In the second line, we divide the domain of integration 
because we want to expand 
$\sqrt{k^2 - (\frac{\sigma}{\kt^{\delta}})^2}$. This is possible 
only when $k \gg \sigma/\kt^{\delta}$, i.e. in the second term.
Now,
\be
\int_{\frac{\sigma}{\kt^{\delta/2}}}^{(\ki\kt)^{\epsilon}} dk\;k^2 G(k)
= \inty dk\;k^2 G(k) 
- \int_{0}^{\frac{\sigma}{\kt^{\delta/2}}} dk\;k^2 G(k)
- \int_{(\ki\kt)^{\epsilon}}^{\infty} dk\;k^2 G(k)\;\;.
\label{cc}
\ee
The second integral in Eq.(\ref{cc}) is of $O(\frac{\sigma}{\kt^{\delta/2}})$,
while the third integral can be neglected since $G(k)$ falls faster than 
any power of $1/k$, \cite{me}. This implies that, for a suitable constant $E$,
\bea
\left|C(a,\ki, \kt)\right| 
&>& \left| \inty dk\;k^2 G(k) \right| - \frac{E\sigma}{\kt^{\delta/2}} 
    \nonumber \\
&>& \half\left|\inty dk\;k^2 G(k)\right|
\label{Ca}
\eea
provided $\inty dk\;k^2 G(k) \neq 0 $. In Appendix \ref{sec:genericity}, 
we show that this is indeed the case for generic $G(k)$. 
Hence, we conclude that the estimate of Eq.(\ref{I2}) is,
\bea
\left|\ki\int_{\kt-\ki}^{\kt+\ki} dk\;kG(k) 
\frac{\theta (\kt, \ki ; k)}{\pi}\right| 
&>& \frac{1}{2\pi} \sqrt{\frac{\ki}{\kt}} 
\left| \inty dk\;k^2 G(k) \right|
- \frac{\ki}{(\ki\kt)^{3/2 - 3\epsilon}} D\;\nonumber\\
&>& \frac{1}{4\pi} \sqrt{\frac{\ki}{\kt}} 
\left| \inty dk\;k^2 G(k) \right|
\eea
for sufficiently large $\ki, \kt$. 

Before we estimate $F(\ki, \kt)$, we need show that Eq.(\ref{I3}) 
is negligible compared to Eq.(\ref{I2}). For the range of variables 
we consider, Eq.(\ref{I3}) satisfies,
\bea
\left| \int_{\kt-\ki}^{\kt+\ki} dk\;kG(k)(\kt - \ki)
\frac{\theta (\kt, k ; \ki)}{\pi}\right|
&<& a\int_{a}^{\ki + \kt} dk\;\left|kG(k)\right| \nonumber \\
&<& a\int_{0}^{\infty} dk\;\left|kG(k)\right| \nonumber \\
&=& O(\frac{1}{\kt^{\delta}})\;\;.
\eea
where we used the fact $ 0 < \theta < \pi$ in the first line.

Now, we are ready to estimate $F(\ki, \kt)$. From the 
Eqs.(\ref{I1a}), (\ref{D}), (\ref{I2b}) and (\ref{Ca}), 
\bea
\left|F(\ki, \kt)\right|
&>& \frac{1}{8\pi} \sqrt{\frac{\ki}{\kt}}\left| \inty dk\;k^2 G(k) 
\right| \nonumber \\
&>& \frac{1}{16\pi}\left| \inty dk\;k^2 G(k) \right| \nonumber \\
&:=& C_{0} \;\;,
\eea
for sufficiently large $\ki, \kt$ and $\kt -\ki \in  
 \left[\frac{1}{\kt^{\delta}}, \frac{2}{\kt^{\delta}}\right]$.

Finally, we have all the necessary ingredients to estimate 
the integral in Eq.(\ref{sinfty}) in the range of interest. 
First, we change variables from $(\ki, \kt)$ to $(a, \kt)$, 
where $a = \kt - \ki$ as before. Then, we integrate $|F|^2$ 
over the domain $a \in \left[\frac{1}{\kt^{\delta}}, 
\frac{2}{\kt^{\delta}}\right]$, and $\kt \in \left[(\kt)_{0}, 
2(\kt)_{0}\right];\; (\kt)_{0} \rightarrow \infty$. 
The integral in Eq.(\ref{sinfty}) is, then,   
\bea
\half\inty d\ki \inty d\kt \left|F(\ki,\kt) \right|^2
&>&\half\int_{(\kt)_{0}}^{2(\kt)_{0}} d\kt 
   \int_{1/\kt^{\delta}}^{2/\kt^{\delta}}da \left|F(\ki,\kt) \right|^2
   \nonumber \\
&>&\half\int_{(\kt)_{0}}^{2(\kt)_{0}} d\kt 
   \int_{1/\kt^{\delta}}^{2/\kt^{\delta}}da (C_{0})^{2} \nonumber \\
&=&\half\frac{(C_{0})^{2}}{1-\delta}(\kt)_{0}^{1-\delta}\left[ 2^{1-\delta} 
 - 1^{1-\delta} \right] \rightarrow \infty
\eea
for small $\delta$. Therefore, the action of the normal-ordered 
generator of generic spatial diffeomorphisms of $\Sigma_{0}$ 
on the Fock vacuum is not well-defined.

\section{Discussion}
\label{sec:conclusion}

The work of Kucha{\v r} \cite{karelcylwave} maps the dynamics of 
cylindrical waves to that of an axisymmetric free scalar field along 
arbitrary axisymmetric foliations of the fixed (2+1)-dimensional  
flat spacetime (\ref{flatmetric}). In this work we studied quantum 
evolution of this free field operator from the initial flat $T=0$ slice, 
$\Sigma_0$, with radial coordinate $R$, to an arbitrary slice $\Sigma$ 
with radial coordinate $r$. We showed that operator evolution is 
unitarily implemented in the standard Fock representation when $r$ 
is chosen to coincide with $R$ on $\Sigma$. The transition to a general 
radial coordinate $r(R)$ from $R$ is obtained by the action of 
a corresponding spatial diffeomorphism on the field variables on $\Sigma$.
We showed that for generic choices of $r(R)$, this diffeomorphism 
is not unitarily implemented in the standard Fock space representation.

In (1+1)-dimensional parametrized field theory (PFT) where such 
operator evolution is unitary, a Tomonaga-Schwinger type of functional 
evolution can be defined as the unitary evolution of Fock states from the 
initial $T=0$ surface to an arbitrary one \cite{charlie1}. 
Our results in this work indicate that a similar notion of 
functional evolution for quantum cylindrical waves is not a viable concept 
for generic choices of foliations starting from the slice $\Sigma_0$. 
One may inquire if such functional evolution can be defined in 
the standard Fock representation by choosing some other slice than 
$\Sigma_0$ as the initial slice (this would entail expressing 
the standard mode coefficients $a(k)$ as functionals of data on 
the new initial slice and analysing the Bogolubov transformation 
corresponding to evolution of these mode coefficients).\footnote{
Note that an affirmative answer for the existence of generic 
evolution from the new initial slice would be consistent with 
our considerations here, only if the specific evolution between 
the new initial slice and $\Sigma_{0}$ was non-unitary.} 
Though we have not investigated this question, we suspect that 
no such choice of initial slice can render generic evolution of 
operators as unitary.

Our results do indicate, however, that it should be possible to 
define functional evolution of quantum states along the restricted 
class of foliations wherein the radial coordinate on each slice is $R$. 
Such evolution is formally described by the `half parametrized' formalism 
of \cite{karelcylwave}. It would be of interest to construct 
the Schr\"odinger picture states as the unitary image of 
the Heisenberg-picture Fock states and to show that they satisfy 
a functional Schr\"odinger equation along the lines of \cite{charlie1}, 
for this restricted evolution.

Our results have been obtained for foliations satisfying the boundary 
conditions (\ref{bct}) and (\ref{bcr}). The fact that $T^{\prime}(r) 
$ and $R(r)-r$ vanish at the axis and at spatial infinity are 
direct consequences of the boundary conditions for cylindrical waves 
in their ADM description (see the appendix of \cite{charliejoe}). 
Our conditions of compact support for these quantities are more 
restrictive than the conditions in \cite{charliejoe}. It would be 
of interest to seek, either a generalization of our proofs to those 
boundary conditions, or an alternative set of boundary conditions 
which allows 
the definition of a consistent Hamiltonian framework and which induce 
conditions on $T(r), R(r)$ such that our proofs still go through. 
We suspect that requiring fall offs of various phase space variables 
to be faster than any power of $r^{-1}$ at spatial infinity and 
any power of $r$ at the axis should suffice for the latter. 
It would be good to check this.

In section V we showed that a generic infinitesimal generator of spatial 
diffeomorphisms on $\Sigma_0$ does not have a well defined action 
on the Fock vacuum. What about generators of infinitesimal `half 
parametrized' evolution on $\Sigma_0$? From \cite{karelcylwave} 
the relevant Hamiltonian density on any slice $\Sigma$ with radial 
coordinate $R$ is given by
\be
{\cal H} = \frac{1}{2}(1-T_{,R}^2)^{-1}
[(R^{-\frac{1}{2}} \pi -R^{\frac{1}{2}}T_{,R} \psi_{,R})^2 
+\frac{1}{2}R \psi_{,R}^2]\;\;.
\ee
On $\Sigma_0$, $T_{,R}=0$, $\pi = R\psi_{,T}$ and $\cal H$ equals 
the standard flat spacetime energy density. $H(f) := \int dR f(R)
{\cal H}$ is a generator of `half parametrized' evolution on $\Sigma_0$. 
In \cite{me} it was shown that for smooth $f$ vanishing fast enough at 
the axis and at spatial infinity, the normal-ordered operator 
${\hat H}(f)$ is densely defined and has a well-defined action on the 
Fock vacuum. Since we know that the standard flat space Hamiltonian, 
${\hat H}_0$, is also a well-defined operator, we conclude from \cite{me} 
that ${\hat H}(f+1) = {\hat H}(f) + {\hat H}_0 $ which generates 
nontrivial evolution at infinity and along the axis is also well-defined. 

It can be checked that the Poisson bracket between two such generators of 
`half parametrized' evolution on $\Sigma_0$, $H(f_1)$ and $H(f_2)$ is the 
generator of a spatial diffeomorphism. It seems puzzling that despite   
${\hat H}(f_1), {\hat H}(f_2)$ being densely defined operators, the generators 
of generic spatial diffeomorphisms do not have a well-defined action on the 
Fock vacuum. A possible resolution is that the relevant shift vector does not 
satisfy our requirements of genericity since it is built out of 
$f_1, f_2$ in a specific way. However, we suspect that the resolution lies 
in the well-known problem of domains for unbounded operators 
(see for e.g. \cite{reedsimon}): although the vacuum is in the 
domain of both ${\hat H}(f_1)$ and ${\hat H}(f_2)$, it need not be so 
for their product and/or commutator. It would be of interest to confirm this.

Next, we comment on the relevance of the recent work of 
Corichi, Cortez, and Mena Marug\'an \cite{unit} to our considerations here. 
They considered the issue of unitary implementability of evolution in 
the context of a quatization of the Gowdy model. The dynamics of the 
Gowdy model is also determined by that of a scalar field on an auxilliary 
2+1 dimensional spacetime. Pierri \cite{monica} showed that the scalar field 
admits a natural quantization based on the choice of a specific foliation 
of the auxilliary spacetime. Corichi et al. \cite{corichi} 
and Torre \cite{torre} 
showed that in the context of this fixed foliation and Pierri's quantization, 
scalar field evolution is not unitarily implemented. However, in  \cite{unit}, 
a quantization was constructed in which the evolution of a time-dependent 
rescaling of the scalar field is unitary. As far as we can tell, 
this quantization is inequivalent to that of \cite{monica}. In view of the 
results of \cite{unit}, the following questions are of interest in the 
context of our work here. Given an arbitrary but fixed axisymmetric 
foliation of 2+1 dimensional flat spacetime, 
is there a quantization in which a suitably redefined scalar field variable 
evolves unitarily? If so, is this quantization equivalent to the standard 
Fock quantization?

We conclude with some remarks regarding the relevance of our results to 
a Dirac quantization of PFT and of gravity. At first glance, 
our results seem to indicate that a straightforward Dirac quantization 
of axisymmetric PFT in 2+1 dimensions (and hence of quantum 
cylindrical waves) which is equivalent to the standard 
Fock quantization, does not exist. \footnote{Note that 
a {\em classical} Hamilton- Jacobi type of canonical transformation 
along the lines of \cite{kareliyer} should result in the 
classical Heisenberg picture constraints which directly yield, 
upon Dirac quantization, the standard Fock space. 
However the PFT constraints which are naturally available 
take the form appropriate to the classical Schr\"odinger picture 
\cite{kareliyer,karel1+1}. Our results imply that 
a direct Dirac quantization of the latter, without further canonical 
transformations, is not equivalent to the standard Fock representation. } 
Indeed it seems that the lack of unitary implementability of 
spatial diffeomorphisms for quantum cylindrical waves has 
an adverse lesson for Loop Quantum Gravity (see for example 
\cite{carlolivingreview}) since the entire formalism is based on 
a unitary representation of spatial diffeomorphisms. 
However, a more careful Loop Quantum Gravity type treatment of PFT 
merits the construction of a kinematic Hilbert space representation 
for both the matter {\em and} the embedding variables rather than 
formally setting the action of ${\hat X^{\alpha}}(x)$ to be 
multiplication by the embedding variable $X^{\alpha}(x)$ and 
that of the conjugate momentum variable to be 
${\hat P_{\alpha}}(x)= -i\frac{\delta}{\delta X^{\alpha}(x)}$. 
Such a treatment is currently in progress \cite{menow} and 
suggests that when the embedding sector is properly defined in 
the quantum theory along with an appropriate notion of embedding 
dependent Hilbert spaces for the matter sector, a satisfactory 
Dirac quantization which is equivalent to the standard Fock quantization 
can be constructed notwithstanding the results of this work and 
\cite{charlie2}.
               

\section*{Acknowledgments}

We thank Charles Torre for useful comments on this work. 
We also thank an anonymous refree for going through the 
manuscript with great care.


\appendix

\section{Lemmas}

\noindent
\begin{Lem}
Consider an integral
\be
I(p,q) = \int_{a}^{b} \; dr g(r)f_{1}(pr) e^{iqf_{2}(r)} e^{ipf_{3}(r)} \;,
\ee
that satisfies 
\begin{itemize}
\item $p,q > 0$.
\item $ g(r), f_{1}(r), f_{2}(r), f_{3}(r) $ are $C^{\infty}$.
\item $ g(r) $ has support only in $[a, b]$.
\item $ f^{\prime}_{2} \neq 0$ in $[a,b]$.
\item $f_{1}$ and all its derivatives are bounded in $(0, \infty) $.
\end{itemize}
Then,
\be
\arrowvert I \arrowvert < \frac{p}{q} M_{1} + \frac{1}{q} M_{2} \;,
\label{aa}
\ee
for suitable constants $M_{1}$ and $M_{2}$.
\label{lemma2}
\end{Lem} 
$\bf{Proof:}$ Let $f_{2}(r) = y$. Since $ f^{\prime}_{2} \neq 0$, 
and $f$ is $C^{\infty}$ in $[a,b]$, $f_{2}$ has a $C^{\infty}$-inverse 
in $[a,b]$, $\chi (y) = r$. Then,
\be
I(p,q) = \int_{\bar{a}}^{\bar{b}}\;dy \left( \frac{d\chi}{dy} \right) g
\left[ \chi(y) \right] f_{1}\left[ p\chi(y) \right] e^{ipf_{3}
\left[ \chi(y)\right]}e^{iqy}\;\;,
\ee
where $\chi(\bar{a}) = a$ and $\chi(\bar{b}) = b$. After integration 
by parts we get Eq.(\ref{aa}) because all functions and their 
derivatives in the integral are $C^{\infty}$ and bounded. 

\noindent
\begin{Lem}
\begin{itemize}
\item $\bf{L.2.a:}$  $\int_{a}^{b} \; dr g(r) e^{ikf(r)} \rightarrow 0$ 
faster than $\frac{1}{k^{n}}$ for any $n$ as $k \rightarrow \infty$, 
if $f^{\prime}(r) \neq 0$ in $[a,b]$, $g(r)$ and $f(r)$ 
are $C^{\infty}$, and $g(r)$ has compact support in $[a,b]$.
\item $\bf{L.2.b:}$ $\int_{a}^{b} \; dr g(r) e^{ikf(r)} \rightarrow 0$ 
as $\frac{1}{k}$ for $k \rightarrow \infty$, if $f^{\prime}(r) \neq 0$ 
in $[a,b]$, $g(r)$ and $f(r)$ are $C^{\infty}$, but either $g(a)$ or 
$g(b)$ or both are non-zero.
\end{itemize}
\end{Lem}
$\bf{Proof:}$ Similar to Lemma $\ref{lemma2}$.


\section{Asymptotic Expansion of Bessel Functions}
\label{sec:asymptotic}
From \cite{watson} p199 we have that for large $\left|z\right|$ and 
$\left|arg\; z\right| < \pi$,
\bea
J_{0}(z) &\sim& \sqrt{\frac{2}{\pi z}} \left[ \cos (z - \frac{\pi}{4}) 
+ \sin (z - \frac{\pi}{4}) \frac{1}{8z} 
- \cos (z - \frac{\pi}{4}) \frac{9}{128z^{2}}
+O(\frac{1}{z^{3}}) \right] \\
J_{1}(z) &\sim& \sqrt{\frac{2}{\pi z}} \left[ \cos (z - \frac{3\pi}{4}) 
- \sin (z - \frac{3\pi}{4}) \frac{3}{8z} 
+ \cos (z - \frac{3\pi}{4}) \frac{15}{128z^{2}}
+ O(\frac{1}{z^{3}}) \right] \;\;.
\eea
\label{asympt}


\section{Derivation of Eq.(\ref{btt}) from Eq.(\ref{bt})}
\label{sec:newapp}

Consider an integral
\bea
I(\ki,\kt, k)&:=& ik\inty dr\;r\jokir\jiktr T^{\prime}
                  e^{ikT(r)}\nonumber\\
             &=& \inty dr\;r\jokir\jiktr
                 \frac{d}{dr}[e^{ikT(r)} -  e^{ikT(\infty)}]\nonumber\\
             &=& \inty dr\;r [\ki\jikir\jiktr - \kt\jokir\joktr]
                 [e^{ikT(r)} -  e^{ikT(\infty)}]\;\;,
\label{appd1}
\eea
where we use integration by parts and Bessel function identities, 
$\frac{d}{dz}\left [ z^{n}J_{n}(z) \right] = z^{n}J_{n-1}(z)$, 
$\frac{d}{dz} J_{0}(z) = - J_{1}(z)$ in the third line. Then, 
\bea
I(\ki,\kt, k) &+& I(\kt,\ki, k)\nonumber\\
&=& (\ki + \kt)\inty dr\; r [\jikir\jiktr - \jokir\joktr]
    e^{ikT(r)}\nonumber\\
&-& (\ki + \kt)\inty dr\; r [\jikir\jiktr - \jokir\joktr]
    e^{ikT(\infty)}\nonumber\\
&=& (\ki + \kt)\inty dr\; r [\jikir\jiktr - \jokir\joktr]
    e^{ikT(r)}\nonumber\\
&-& (\ki + \kt)e^{ikT(\infty)}\left[\frac{\delta(\ki,\kt)}{\ki} 
    -\frac{\delta(\ki,\kt)}{\ki}\right]\nonumber\\
&=& (\ki + \kt)\inty dr\; r [\jikir\jiktr - \jokir\joktr]
    e^{ikT(r)}\;\;.
\label{appd2}
\eea
Here we use an orthogonality relation for Bessel functions,
\be
\inty dr\;r J_{n}(k r)J_{n}(l r) = \frac{\delta(k, l)}{k}\;\;,
\label{orthogonal}
\ee
in the fifth line. Eq.(\ref{bt}) is then,
\bea
\beta(\ki, \kt) &=& -\frac{\ki}{2(\ki + \kt)}
                     \left[ I(\ki,\kt,\kt) + I(\kt,\ki,\kt)\right]\nonumber\\
                &=& -\frac{i}{2}\frac{\ki\kt}{(\ki + \kt)} 
                     \int_{a}^{b} dr\;r [\jokir\jiktr 
                     +\jikir\joktr]\;T' e^{i\kt T} \;\;,
\eea
which is Eq.(\ref{btt}). In the last line, we use the compactness 
of $T^{\prime}$, Eq.(\ref{bct}).

\section{Derivation of Eq.(\ref{hankelF})}
\label{sec:appB}

In this section we present the derivation of Eq.((\ref{hankelF}). 
The Hankel transform $G(k)$ of $rf(r)$ with respect to $\jokr$ is 
[See, \cite{watson} p453]
\be
G(k) = \inty \;dr \;r\; rf(r) \jokr 
\label{g}
\ee
\be
rf(r)  = \inty \;dk\;k\; G(k)\jokr 
\label{hankel}
\ee
Using Eq.(\ref{hankel}) into Eq.(\ref{F}) we get
\be
F(\ki\kt) = \ki\kt\inty \; dk G(k)k \inty dr 
\left[\jokr\jokir\jiktr + \jokr\jikir\joktr \right].
\label{f}
\ee
From p411 of \cite{watson}
\bea
\inty dr \jokr\jokir\jiktr 
&=& \frac{\theta (\ki, k ; \kt)}{\pi\kt}  \;\;\;\; 
    \text{if} \;\;\left|k-\ki \right| 
<   \kt <\left| k + \ki \right| \nonumber \\
&=& \frac{1}{\kt} \;\;\;\;\;\;\;\;\;\;\;\;\;\;\;\;\;      
    \text{if} \;\;\kt > k + \ki \nonumber \\
&=& 0\;\;\;\;\;\;\;\;\;\;\;\;\;\;\;\;\;\;\;                    
    \text{otherwise}                                     
\label{hankel1}
\eea
and
\bea 
\inty dr \jokr\joktr\jikir 
&=& \frac{\theta (\kt, k ; \ki)}{\pi\ki}  \;\;\;\; 
    \text{if} \left| k-\kt \right| 
<   \ki < \left| k + \kt \right| \nonumber \\
&=& \frac{1}{\ki} \;\;\;\;\;\;\;\;\;\;\;\;\;\;\;\;\;       
    \text{if} \;\;\ki > k + \kt \nonumber \\
&=& 0\;\;\;\;\;\;\;\;\;\;\;\;\;\;\;\;\;\;\;                     
    \text{otherwise} \;\;.                                    
\label{hankel2}
\eea
Using Eq.(\ref{hankel1}) and Eq.(\ref{hankel2}) into Eq.(\ref{f}) 
we get Eq.(\ref{hankelF}).


\section {Proof of $\inty dk k^2 G(k) \neq 0$ for generic $G(k)$ }
\label{sec:genericity}

Recall, from Eq.(\ref{g}) and Eq.(\ref{hankel}), that $G(k)$ 
is the Hankel transform of $G(r) := rf(r)$ with respect to 
$\jokr$. Using an integral representation of $\jokr$ and 
rotational symmetry of $G(r)$, $G(\vec{k}) 
:= G(k)$ can be shown to be a two-dimensional Fourier transform of 
$G(\vec{r}) := G(r)$\cite{me}. That is,
\be
G(\vec{k}) = \frac{1}{2\pi}\int dx^2 G(x_{1}, x_{2}) 
e^{i\ki x_{1} + i\kt x_{2}}
\ee
However we can also consider $G(\ki, \kt = 0)$ as a one-dimensional 
Fourier transform of $g(x_{1})$ as defined below,
\bea
G(\ki, \kt = 0) 
&=& \frac{1}{\sqrt{2\pi}}\int_{-\infty}^{\infty}dx_{1}
    e^{i\ki x_{1}}\left[ \frac{1}{\sqrt{2\pi}}
    \int_{-\infty}^{\infty}dx_{2} G(x_{1}, x_{2})\right]\nonumber\\
&:=&\frac{1}{\sqrt{2\pi}}\int_{-\infty}^{\infty}dx_{1}
    e^{i\ki x_{1}}g(x_{1})\;\;.
\label{G}
\eea
Note that $g(x_{1})$ exists and posses good fall-off and smoothness 
property inherited from $G(x_{1}, x_{2})$. Then, $G(\ki, \kt = 0) 
:= g(\ki)$ satisfies,
\be
g(x_{1})=\frac{1}{\sqrt{2\pi}}\int_{-\infty}^{\infty}d\ki
         e^{-i\ki x_{1}}g(\ki)\;\;. 
\ee
Now,
\be
\frac{d^2 g(x_{1})}{dx_{1}^{2}} = -\frac{1}{\sqrt{2\pi}}
                                  \int_{-\infty}^{\infty}d\ki
                                  e^{i\ki x_{1}}\ki^2 g(\ki)\;\;,
\ee
which implies
\bea
\left.\frac{d^2 g(x_{1})}{dx_{1}^{2}}\right|_{x_{1}=0} 
&=& -\frac{1}{\sqrt{2\pi}}\int_{-\infty}^{\infty}d\ki \ki^2 g(\ki)\nonumber\\
&=& -\sqrt{\frac{2}{2\pi}}\inty d\ki \ki^2 g(\ki)\nonumber\\
&=& -\sqrt{\frac{2}{2\pi}}\inty dk k^2 G(k)\;\;.
\eea
Here we have used the fact that $G(\ki, \kt=0) := g(\ki)$, 
and $G(\vec{k}) = G(\|\vec{k}\|)$ due to the rotational symmetry.

To prove $\inty dk k^2 G(k) \neq 0$ for generic $G(k)$ 
we only need to prove that $\left.\frac{d^2 g(x_{1})}{dx_{1}^{2}}
\right|_{x_{1}=0} \neq 0$ generically. For fixed $x_{1}$, we change 
variables from $(x_{1}, x_{2})$ to $(x_{1}, r)$; $x_{2} 
\mapsto \sqrt{x_{1}^2 + x_{2}^2} = r$. 
Then, from Eq.(\ref{G}) and rotational symmetry of $G(\vec{x})$,
\bea
g(x_{1})
&=& \sqrt{\frac{2}{\pi}}\inty dx_{2} G(x_{1}, x_{2}) \nonumber \\
&=& \sqrt{\frac{2}{\pi}}\int_{x_{1}}^{\infty} dr r \frac{G(r)}
    {\sqrt{r^2 - x_{1}^2}}\;\;.
\eea
Since $G(r)$ is a function with compact support away from $r=0$, 
there exists a constant $r_{0}$ such that $G(r) = 0$ for $r < r_{0}$.
Then, for $x_{1} < r_{0}$, we get,
\be
\sqrt{\frac{2}{\pi}}\int_{x_{1}}^{\infty} dr r \frac{G(r)}
    {\sqrt{r^2 - x_{1}^2}}
= \sqrt{\frac{2}{\pi}}\int_{r_{0}}^{\infty} dr r \frac{G(r)}
    {\sqrt{r^2 - x_{1}^2}}\;\;.
\ee
That is, for $x_{1} < r_{0}$,
\be
\frac{d^2 g(x_{1})}{dx_{1}^{2}} 
= \sqrt{\frac{2}{\pi}}\int_{r_{0}}^{\infty} dr r G(r)
    \frac{d^2}{dx_{1}^{2}}
    \left(\frac{1}{\sqrt{r^2 - x_{1}^2}}\right)\;\;.
\label{dtrick}
\ee
Since we are interested in the limit $x_{1} \rightarrow 0$, we can 
use Eq.(\ref{dtrick}). Then, 
\bea
\left.\frac{d^2 g(x_{1})}{dx_{1}^{2}}\right|_{x_{1}=0}\nonumber 
&=& \left.\sqrt{\frac{2}{\pi}}\int_{r_{0}}^{\infty} dr r G(r)
    \frac{d^2}{dx_{1}^{2}}\right|_{x_{1}=0}
    \left(\frac{1}{\sqrt{r^2 - x_{1}^2}}\right) \nonumber\\ 
&=& \sqrt{\frac{2}{\pi}}\int_{r_{0}}^{\infty} dr\frac{G(r)}{r^2}\nonumber\\
&=& \sqrt{\frac{2}{\pi}}\int_{r_{0}}^{\infty} dr \frac{f(r)}{r}\nonumber\\
&=& \sqrt{\frac{2}{\pi}}\int_{0}^{\infty} dr \frac{f(r)}{r}\;\;,
\eea
where we use the fact $f(r)=0$ for $r<r_{0}$ in the last line. 
For generic embedding the integral in the last line does not vanish, 
therefore, $\inty dk k^2 G(k) \neq 0$. 



\end{document}